\documentclass[preprint,showpacs,preprintnumbers,amsmath,amssymb,floatfix]{revtex4}

\usepackage{graphicx}% Include figure files
\usepackage{dcolumn}% Align table columns on decimal point
\usepackage{bm}% bold math

\def\lcdm  {\rm $\Lambda$CDM\ }

\def\kpc{{\rm\,kpc}}

\def\mpc{{\rm\,Mpc}}
\def\ihmpcC{h^3 {\rm\,Mpc}^{-3}}

\def\kms{{\rm\,km/s}}

\def\vol#1  {{{#1}{\rm,}\ }}

\def\etal{et al.\ }

\newcount\refno
\newcount\rfno
\def\eq{$^{\the\refno\ }$\advance\refno by 1}
\def\ad{\advance\rfno by 1}

\def\clock{\count0=\time \divide\count0 by 60
     \count1=\count0 \multiply\count1 by -60 \advance\count1 by \time
     \number\count0:\ifnum\count1<10{0\number\count1}\else\number\count1\fi}

\def\myputfigure#1#2#3#4#5%
{\hskip0.03\textwidth\vskip#5pt%\hfill
\makebox[0pt]{\hskip#2in
\includegraphics[width=#3\textwidth]{#1}}\vskip#4pt\hfill}

\def\kms{\rm km\,s^{-1}}

\def\ergs{\rm erg\,s^{-1}}
\def\Gcm2{\rm G~cm^2}

\def\beq{\begin{equation}}
\def\eeq{\end{equation}}
\def\bea{\begin{eqnarray}}
\def\eea{\end{eqnarray}}

\def \eg           {{e.g.}}
\def \date         {\ifcase\month \message{zero} \or
                    January \or February \or March \or April \or May \or June
                    \or July \or
                    August \or September \or October \or November \or
                    December \fi
                    \space\number\day, \number\year}

\def\kev{{\rm keV}}

\def\cmb{_{{\rm CMB}}}

\def\bright{{\rm bright}}
\def\faint{{\rm faint}}
\def\total{{\rm total}}

\def\lang{\langle}
\def\rang{\rangle}

\begin{document}

\newcommand\hatbfn{{\bf\hat{n}}}

\preprint{v2.1, astro-ph/0308260}

\title{Cross-Correlation of the Cosmic Microwave Background with the 2MASS Galaxy
Survey:\\
Signatures of Dark Energy, Hot Gas, and Point Sources}

\author{Niayesh Afshordi}
 \email{afshordi@astro.princeton.edu}
 \affiliation{Princeton University Observatory, Princeton, NJ 08544, USA}%Lines break automatically or can be forced with \\
\author{Yeong-Shang Loh}
 \affiliation{Princeton University Observatory, Princeton, NJ 08544, USA}
%\author{David N. Spergel}
 %\affiliation{Princeton University Observatory, Princeton, NJ 08544, USA}
\author{Michael A. Strauss}
 \affiliation{Princeton University Observatory, Princeton, NJ 08544, USA}
% \textbackslash\textbackslash

\date{}% It is always \today, today,

\begin{abstract}
We cross-correlate the Cosmic Microwave Background (CMB)
temperature anisotropies observed by the
   Wilkinson Microwave Anisotropy Probe (WMAP) with the projected distribution of extended sources in the
   Two Micron All Sky Survey (2MASS). By modelling the theoretical
   expectation for this signal, we extract the signatures of dark energy (Integrated Sachs-Wolfe effect;ISW),
   hot gas (thermal Sunyaev-Zeldovich effect;thermal SZ), and microwave point sources in the cross-correlation.
    Our strongest signal is the thermal SZ, at the $3.1-3.7\sigma$ level, which is consistent with
    the theoretical prediction based on observations of X-ray clusters. We also see the ISW signal at the
$2.5\sigma$ level, which
    is consistent with the expected value for the concordance \lcdm cosmology, and is an independent signature of the
    presence of dark energy in the universe. Finally, we see
    the signature of microwave point sources at the $2.7\sigma$ level.

\end{abstract}

%\pacs{98.65.Dx,98.62.Py,98.70.Vc,98.80.Es}
\maketitle

\section{Introduction}

The recently released WMAP \citep{BennettEtAl2003a} results
constrain our cosmology with an unprecedented accuracy. Most of
these constraints come from the linear fossils of the early
universe which have been preserved in the temperature anisotropies
of the CMB. These are the ones that can be easily understood and
dealt with, within the framework of linear perturbation theory.
However, there are also imprints of the late universe which could
be seen in the WMAP results. Most notably, the measurement of the
optical depth to the surface of last scattering, $\tau\simeq
0.17$, which implied an early reionization of the universe, was
the biggest surprise. There is also the strangely small amplitude
of the large-angle CMB anisotropies which remains
unexplained\citep{SpergelEtAl2003}.

Can we extract more about the late universe from WMAP? Various
secondary effects have been studied in the literature (see \eg
\citep{HuDodelson2002} which lists a few). The main secondary
anisotropy at large angles is the so-called Integrated Sachs-Wolfe
(ISW) effect\citep{SachsWolfe1967}, which is a signature of the
decay of the gravitational potential at large scales. This could
be either a result of spatial curvature, or presence of a
component with negative pressure, the so-called dark energy, in
the universe\citep{ratra}. Since WMAP has constrained the
deviation from flatness to less than 4\%, the ISW effect may be
interpreted as a signature of dark energy. At smaller angles, the
dominant source of secondary anisotropy is the thermal
Sunyaev-Zeldovich (SZ) effect\citep{SZ}, which is due to
scattering of CMB photons by hot gas in the universe.

However, none of these effects can make a significant contribution
to the CMB power spectrum below $\ell\sim 1000$, and thus they are
undetectable by WMAP alone. One possible avenue is
cross-correlating CMB anisotropies with a tracer of the density in
the late universe\citep{Turok,PeirisSpergel2000,Cooray2002}.
 This was first done by \citep{Bou02} who cross-correlated the
COBE/DMR map\citep{cobe} with the NRAO VLA Sky Survey
(NVSS)\citep{nvss}. After the release of WMAP, different groups
cross-correlated the WMAP temperature maps with various tracers of
the low-redshift universe. This was first done with the ROSAT
X-ray map in \citep{diego}, where a non-detection of the thermal
SZ effect puts a constraint on the hot gas content of the
universe. \citep{hernand} claimed a 2-5$\sigma$ detection of an SZ
signal by filtering WMAP maps via templates made using known X-ray
cluster catalogs. \citep{Nol03} looked at the cross-correlation
with the NVSS radio galaxy survey, while
\citep{boughn/crittenden:2003} repeated the exercise for NVSS, as
well as the HEAO-1 hard X-ray background survey, both of which
trace the universe around redshift of $\sim 1$. Both groups found
their result to be consistent with the expected ISW signal for the
WMAP concordance cosmology \citep{BennettEtAl2003a}, i.e. a flat
$\Lambda$CDM universe with $\Omega_m \simeq 0.3$, at the
$2-3\sigma$ level. Their result is consistent with the \lcdm
paradigm which puts most of the energy of the universe in dark
energy\citep{ratra}.

  More recently, \citep{fosalba1} cross-correlated WMAP with the
  APM galaxy survey\citep{apm} which traces the galaxy distribution at $z\sim 0.15$.
  This led to an apparent detection of both the thermal SZ and ISW signals. However,
  the fact that they use a jack-knife covariance matrix to estimate
  the strength of their signal, while their jack-knife errors are
  significantly
  smaller than those obtained by Monte-Carlo realizations of the CMB
  sky (compare their Figure 2 and Figure 3) weakens the significance of their claim.
  Indeed, as we argue below (see III.C), using Monte-Carlo realizations of
  the CMB sky is the only reliable way to estimate a covariance
  matrix if the survey does not cover the whole sky.

   \citep{Myers2003} cross-correlates the highest frequency band (W-band) of
   the WMAP with the ACO cluster survey\citep{ACO}, as well as the galaxy groups and clusters in the
   APM galaxy survey. They claim a $2.6\sigma$ detection of
   temperature decrement on angles less than $0.5^\circ$,
   which they associate with the thermal SZ effect. However they
   only consider the Poisson noise in their cluster distribution
   as their source of error. This underestimates the error due to
   large spatial correlations (or cosmic variance) in the cluster distribution (see III.C).
   \citep{Myers2003} also studies the cross-correlation of the W-band with
   the NVSS radio sources below a degree and claims a positive
   correlation at the scale of the W-band resolution. This may imply a
   possible contamination of the ISW signal detection in
   \citep{boughn/crittenden:2003} and \citep{Nol03}. However the
   achromatic nature of this correlation makes this unlikely\citep{Steve}.

  Finally, \citep{scranton} and \citep{fosalba2} repeated the cross-correlation analysis
  with the 3400 and 2000 square degrees, respectively, of the Sloan
  Digital Sky Survey\citep{YorkEtAl2000}. Both groups claim detection of a positive
  signal, but they both suffer from the inconsistency of their
  jack-knife and Monte-Carlo errors.
  % and \cite{scranton} sees systematic
 % deviations from Gaussian statistics, indicating systematics in error estimates.

   The 2MASS Extended Source Catalog (XSC)\citep{Jar00} is a full sky, near
   infrared survey of galaxies whose median redshift is around $z\sim
   0.1$. The survey has reliable and uniform photometry of about 1 million galaxies, and is
   complete, at the 90\% level for K-magnitudes
   brighter than 14, over $\sim 70\%$ of the sky. The large area coverage
   and number of galaxies makes the 2MASS XSC a good tracer of
   the ISW and SZ signals in the cross-correlation with the CMB.

  In this paper, we study the cross-correlation of the WMAP Q,V and W
  bands with four different K-magnitude bins of the 2MASS Extended
  Source Catalog, and fit it with a three component model which
  includes the ISW, thermal SZ effects and microwave sources. We compare
  our findings with the theoretical expectations from the
  WMAP+CBI+ACBAR+2dF+Ly-$\alpha$
  best fit cosmological model (WMAP concordance model from here on; see Table 3
  in \citep{BennettEtAl2003a}), which is a flat universe with,
  $\Omega_m =0.27, \Omega_b  =0.044, h=0.71$, and $\sigma_8 = 0.84$. We also
assume their values of $n_s=0.93$, and $dn_s/d\ln k = -0.031$ for
the spectral index and its running at $k=0.05 \mpc$
\citep{SpergelEtAl2003}.

     We briefly describe the relevant secondary
     anisotropies of the CMB in Sec. II. Sec. III describes the properties
     of the cross-correlation of two random fields, projected on
     the sky. Sec. IV summarizes the relevant information on
     the WMAP temperature maps and
     the 2MASS Extended Source Catalog. Sec. V describes our
     results and possible systematics, and Sec. VI
     concludes the paper.

\section{What are the secondary anisotropies?}

   The dominant nature of the Cosmic Microwave
   Background (CMB) fluctuations, at angles larger than $\sim 0.1$ degree,
   is primordial, which makes CMB a snapshot of the universe
   at radiation-matter decoupling, around redshift of $\sim 1000$.
   However, a small part of these fluctuations can be generated as the
   photons travel through the low redshift universe.
   These are the so-called secondary anisotropies. In this
   section, we go through the three effects which should dominate
   the
    WMAP/2MASS cross-correlation.
   \subsection{Integrated Sachs-Wolfe effect}
   The first one is the Integrated Sachs-Wolfe (ISW)
  effect\citep{SachsWolfe1967}
  which is caused by the time variation in the cosmic gravitational
  potential, $\Phi$. For a flat universe, the anisotropy
  due to the ISW effect is an integral over the conformal time
  $\eta$
  \bea
  \delta_{\rm ISW}({\bf \hat{n}}) &=& 2 \int \Phi^{'}[(\eta_0-\eta){\bf \hat{n}}, \eta] ~d\eta ,
  \eea
  where $\Phi^{'} \equiv \partial \Phi/\partial \eta$, and ${\bf
  \hat{n}}$ is unit vector in the line of sight.
  The linear metric is assumed to be
  \beq
  ds^2=a^2(\eta)\{[1+2\Phi({\bf x},\eta)]d\eta^2-[1-2\Phi({\bf x},\eta)]{\bf dx \cdot dx} \},
  \eeq
  and $\eta_0$ is the conformal time
  at the present.

  In a flat universe, the gravitational potential $\Phi$ is constant for a
  fixed equation of state and therefore
  observation of an ISW effect is an indicator of a change in the equation
  of state of the universe. Assuming that this change is due to an
  extra component in the matter content of the universe, the so-called dark energy,
  this component should have a negative pressure to become important at
  late times\citep{ratra}. Therefore, observation of an ISW effect in a flat universe is a
  signature of dark energy.

  The ISW effect
  is observed at large angular scales because most of the power in
 the fluctuations of $\Phi$ is at large scales. Additionally, the
 fluctuations at
  small angles tend to cancel out due to the integration over the
  line of sight.

\subsection{Thermal Sunyaev-Zeldovich effect}
  The other significant source of secondary anisotropies is the
  so-called thermal Sunyaev-Zeldovich (SZ) effect \citep{SZ},
   which is caused by the scattering of CMB photons off the hot electrons of the
  intra-cluster medium. This secondary anisotropy is frequency
  dependent, i.e. it cannot be associated with a single change in
  temperature. If we define a frequency dependent $T(\nu)$ so that
  $I_B[\nu;T(\nu)] = I(\nu)$, where $I(\nu)$ is the CMB
  intensity and $I_B[\nu;T]$ is the black-body spectrum at temperature $T$,
  the SZ anisotropy takes the form
  \bea
  \frac{\delta T(\nu)}{T(\nu)}  &=& -
  \frac{\sigma_T f(x)}{m_e c} \int \delta p_e[(\eta_0-\eta){\bf\hat{n}},\eta] a(\eta)
  ~d\eta,
  \eea
  where
  \beq
  x\equiv h\nu/(k_B T_{\cmb}) \,~{\rm and}~\, f(x) \,\,\equiv\,\, 4 -
  x\coth(x/2),
  \eeq
  and $p_e$ is the electron pressure.
  Assuming a linear pressure bias with respect to the matter
  overdensity $\delta_m$:
  \bea
  \frac{\delta p_e}{p_e} &=& b_p \delta_m,
  \eea
  Eq.(3) can be written as
  \beq
  \delta_{\rm SZ}(\nu) \equiv \frac{\delta T(\nu)}{T(\nu)} \,=\, - F(x)\int \tilde{T}_e \delta_m \frac{H_0
  d\eta}{a^2(\eta)},
  \eeq
  where
  \bea
  &\tilde{T}_e = b_p \bar{T}_e,& \nonumber\\
  &F(x)=\frac{n_e k_B \sigma_T f(x)}{4 m_e c H_0} =(1.16 \times
  10^{-4} \kev^{-1}) \Omega_b h f(x),&
  \eea
  and $\bar{T}_e$ and $n_e$ are the average temperature and
  the comoving density of (all) electrons, respectively.
  In Appendix A, we make an analytic estimate for $\tilde{T}_e$, based on the mass function
  and mass-temperature relation of galaxy clusters.

\subsection{Microwave Sources}

Although technically they are not secondary anisotropies,
microwave sources may contribute to the cross-correlation signal,
as they are potentially observable by both WMAP and 2MASS. For
simplicity, we associate an average microwave luminosity with all
2MASS sources. We can relax this assumption by taking this
luminosity to be a free parameter for each magnitude bin, and/or
removing the clustering of the point sources. As we discuss in
Sec. V.C, neither of these change our results significantly.

For the microwave spectrum in different WMAP frequencies, we try
both a steeply falling antenna temperature $\propto 1/\nu^{2-3} $
(consistent with WMAP point sources\citep{mapfor}) and a Milky Way
type spectrum which we obtain from the WMAP observations of the
Galactic foreground \citep{mapfor}.

In Appendix B, assuming an exponential surface emissivity with a
scale length of $5 \kpc$ for the Galactic disk and a small disk
thickness, we use the Galactic latitude dependence of the WMAP
temperature to determine the luminosity of the Milky Way (Eq.B6)
in different WMAP bands:
$$L^*_Q = 1.7 \times 10^{37} ~\ergs,$$
$$L^*_V = 3.0 \times 10^{37} ~\ergs,$$
$$ {\rm and}~ L^*_W = 1.0 \times 10^{38} ~\ergs. \eqno(B6) $$

 These values are within 50\% of the observed WMAP luminosity of the Andromeda galaxy(see Appendix B)
 \footnote{We thank Doug Finkbeiner for extracting the fluxes of Andromeda in WMAP temperature maps.}.
   In V.C, we compare the observed average luminosity of the 2MASS sources to these numbers (see Table II).

   The contribution to the CMB anisotropy due to Point Sources (see Eq.B2) is given by
\beq
   \delta_{PS}({\bf \hat{n}}) = \frac{\delta T({\bf \hat{n}})}{T} = \frac{4\pi^2 \hbar^3 c^2\sinh^2(x/2) L(x)}{(x k_B T_{\cmb})^4 \Delta
   x} \int dr \left(\frac{r}{d_L(r)}\right)^2 n_c(r) \delta_g(r,{\bf \hat{n}}),
\eeq
   where $\Delta x$ is the effective bandwidth of the WMAP band\citep{BennettEtAl2003a}, $n_c(r)$ is
   the average comoving number density of the survey galaxies, $d_L$ is luminosity distance, and
   $\delta_g$ is the galaxy overdensity.

\section{The cross-correlation power spectrum}
\subsection{The Expected Signal}
    We first develop the theoretical expectation
    value of the cross-correlation of two random fields,
    projected on the sky. Let us consider two random
    fields $A({\bf x})$ and $B({\bf x})$ with their Fourier transforms defined
    as
    \beq
    A_{\bf k} = \int {\bf d^3 x}~  e^{-i{\bf k.x}} A({\bf x}), ~ {\rm and} ~B_{\bf k} = \int {\bf d^3 x}~  e^{-i{\bf k.x}}
    B({\bf x}).
    \eeq
      The cross-correlation power spectrum, $P_{AB}(k)$ is defined
      by
    \bea
     \langle A_{\bf k_1} B_{\bf k_2}\rangle &=& (2\pi)^3 {\bf \delta^3(k_1-k_2)} P_{AB}(k_1).
    \eea
    The projections of $A$ and $B$ on the sky are defined using
    $F_A$ and $F_B$ projection kernels
    \beq
    \tilde{A}({\bf \hat{n}}) \,=\, \int dr~F_A(r) A(r{\bf \hat{n}}),\,~{\rm and}~\,
    \tilde{B}({\bf \hat{n}}) \,=\, \int dr~F_B(r) B(r{\bf \hat{n}}).
    \eeq

    For the secondary temperature anisotropies, these kernels were given in
    Eqs.1,6 and 8. For the projected galaxy overdensity, this
    kernel is
    \beq
    F_g(r) =\frac{ r^2~ n_c(r)}{\int dr^\prime ~{r^\prime}^2 ~n_c(r^\prime)}.
    \eeq
    For our treatment, we assume a constant galaxy bias, $b_g$,
    which relates the galaxy fluctuations, $\delta_g$, to the overall matter
    density fluctuations $\delta_m$, up to a shot noise $\delta_p$
    \beq
   \delta_g = b_g \delta_m +\delta_p.
   \eeq
   In this work, we constrain the galaxy bias, $b_g$,
   by comparing the auto-correlation of the galaxies with the
   expected matter auto-correlation in our cosmological model. Our
   bias, therefore, is model dependent.

    Now, expanding $\tilde{A}$ and $\tilde{B}$ in terms of spherical harmonics, the cross-power spectrum,
    $C_{AB}(\ell)$ is defined as
    \bea
    &C_{AB}(\ell)\equiv \langle\tilde{A}_{\ell m}\tilde{B}^*_{\ell m}\rangle&\nonumber\\
             &= \int dr_1 dr_2 F_A(r_1) F_B(r_2) \times &\nonumber\\
             &\int \frac{{\bf d^3
                 k}}{(2\pi)^3} P_{AB}(k) (4\pi)^2 j_{\ell}(kr_1)j_{\ell}(kr_2)
                 Y_{\ell m}({\bf \hat{k}})Y^*_{\ell m}({\bf \hat{k}})&\nonumber\\
             &= \int dr_1 dr_2 F_A(r_1) F_B(r_2) \int \frac{2k^2
                 dk}{\pi}j_{\ell}(kr_1)j_{\ell}(kr_2) P_{AB}(k),&\nonumber\\
                 ~~
    \eea
    where $j_{\ell}$'s are the spherical Bessel functions of rank $\ell$
    and $Y_{\ell m}$'s are the spherical harmonics.

     To proceed further, we use the small angle (large $\ell$)
     approximation for the spherical Bessel functions
    \beq
    j_{\ell}(x)= \sqrt{\frac{\pi}{2\ell+1}}[\delta_{\rm Dirac}(\ell+\frac{1}{2}-x)+{\rm
    O}(\ell^{-2})],
    \eeq
    which yields
    \beq
    C_{AB}(\ell)=\int \frac{dr}{r^{2}} F_A(r)F_B(r) P_{AB}\left(\frac{\ell+1/2}{r}\right) \cdot [1+{\rm
    O}(\ell^{-2})].
    \eeq
    This is the so called Limber equation \citep{limber}. As we do
    not use the quadrupole due to its large Galactic contamination,
    the smallest value of $\ell$ that we use is 3. Direct
    integration of Eq.(15) (for the ISW signal which is dominant for
    low $\ell$'s, see Figure 7) shows that the Limber equation
    overestimates the cross-power by less than 2-3\% at $\ell=3$, which is
    negligible compared to the minimum cosmic variance error (about $40\%$, see III.B) at
    this multipole. Therefore, the Limber equation is an accurate estimator of
    the theoretical power spectrum.

     Now we can substitute the results of Sec. II (Eqs.1, 6, 8 and 12) into Eq.(16) which
     yields
     \bea
     C_{gT}(x,\ell)= \frac{b_g}{\int dr~r^2 n_c(r)} \int dr~
     n_c(r)\{2 P_{\Phi^{\prime},m}\left(\frac{\ell+1/2}{r}\right)\nonumber\\-\left[F(x)\tilde{T}_e
     H_0 (1+z)^2-
     \frac{4\pi^2 \hbar^3 c^2\sinh^2(x/2) b_g L(x)}{(x k_B T_{\cmb})^4 \Delta x
     (1+z)^2}\right ]P\left(\frac{\ell+1/2}{r}\right)\},
     \eea
       where $P(k)$ is the matter power spectrum, $z$ is the
       redshift, and $x$ is defined in Eq.(4). The terms in Eq.(17) are the ISW, SZ and Point Source contributions
       respectively. Since the ISW effect is only important at large scales,
     the cross-power of the gravitational potential derivative with matter
     fluctuations can be expressed in terms of the matter power spectrum, using the Poisson equation and
     linear perturbation theory, and thus we end up with
     \bea
     C_{gT}(x,\ell)= \frac{b_g}{\int dr~r^2 n_c(r)} \int dr~
     n_c(r)\{-3H^2_0 \Omega_m \frac{r^2}{(\ell+1/2)^2} \cdot \frac{g^{\prime}}{g} (1+z) \nonumber\\-F(x)\tilde{T}_e
     (1+z)^2+
     \frac{4\pi^2 \hbar^3 c^2\sinh^2(x/2) b_g L(x)}{(x k_B T_{\cmb})^4 \Delta x
     (1+z)^2}\}P\left(\frac{\ell+1/2}{r}\right),
     \eea
      where $g$ is the linear growth factor of the gravitational
      potential,$\Phi$, and $g^{\prime}$ is its derivative with
      respect to the conformal time. We will fit this model to our
      data in Sec. V, allowing a free normalization for each term.

      Finally, we write the theoretical expectation for the
      projected galaxy auto-power,$C_{gg}$, which we use to find the galaxy
      bias. Combining Eqs.12,13 and 16, we arrive at
      \beq
      C_{gg}(\ell) =\frac{\int dr~r^2~ n^2_c(r) [b^2_g \cdot
      P\left(\frac{\ell+1/2}{r}\right)+ \gamma \cdot n^{-1}_c(r)]}{\left[\int
      dr~ r^2~n_c(r)\right]^2},
      \eeq
      where the $n_c^{-1}$ term is the power spectrum of the Poisson noise, $\delta_p$, while
      the extra free parameter, $\gamma$, is introduced to include
      the possible corrections to the Poisson noise due to the finite pixel
      size. In the absence of such corrections $\gamma=1$. In Sec. V, we seek the values of $b_g$ and $\gamma$ that
      best fit our observed auto-power for each galaxy sub-sample.

     To include the effects of non-linearities in the galaxy power spectrum,
     we use the Peacock \& Dodds fitting formula \cite{peacock}
     for the non-linear matter power spectrum, $P(k)$.

\subsection{Theoretical errors: cosmic variance vs. shot noise}

    To estimate the expected theoretical error, again for simplicity, we restrict the
    calculation to the small angle limit. In this limit, the
    cross-correlation function can be approximated by
    \bea
    C_{AB}(\ell) &\simeq& \frac{4\pi}{\Delta\Omega}\lang\tilde{A}_{\ell m}~\tilde{B}^*_{\ell m}\rang,
    \eea
    where $\Delta \Omega$ is the common solid angle of the patch of the sky covered by observations of both $\tilde{A}$
    and $\tilde{B}$\citep{tegmark}.

     Assuming gaussianity, the standard deviation in $C_{AB}$, for a single harmonic mode, is given
     by
    \bea
    \Delta C^2_{AB}(\ell) &=& \lang C^2_{AB}(\ell)\rang\,-\,\lang C_{AB}(\ell)\rang^2 \nonumber\\
    &=&\Delta \Omega^{-2}[\lang\tilde{A}_{\ell m}~\tilde{B}^*_{\ell m}\rang
                          \lang\tilde{A}_{\ell m}~\tilde{B}^*_{\ell m}\rang
                     \,+\,\lang\tilde{A}_{\ell m}~\tilde{A}^*_{\ell m}\rang
                          \lang\tilde{B}_{\ell m}~\tilde{B}^*_{\ell m}\rang] \nonumber\\
    &=& C^2_{AB}(\ell) \,+\, C_{AA}(\ell)C_{BB}(\ell).
    \eea
    The number of modes available between $\ell$ and $\ell+1$, in the
    patch $\Delta \Omega$, is
    \beq
    \Delta N \simeq \frac{(2\ell+1) \Delta \Omega}{4\pi},
    \eeq
    and so the standard deviation of $C_{AB}$, averaged over all
    these modes is
    \beq
     \Delta C^2_{AB}(\ell) \simeq \frac{4\pi}{ \Delta \Omega(2\ell+1)}
     [C^2_{AB}(\ell)+C_{AA}(\ell)C_{BB}(\ell)].
    \eeq

    In fact, since the main part of CMB fluctuations is of primordial origin, the first term in
    brackets is negligible for the cross-correlation error, so the error in the cross-correlation
    function, as one may expect, depends on the individual
    auto-correlations.

     We can use the CMBFAST code
    \citep{cmbfast} to calculate the auto-correlation of the CMB
    temperature fluctuations. Also, the theoretical expectation for the auto-power of the
    projected galaxy distribution is given by Eq.(19).

      The galaxy/CMB auto-power spectra are dominated by Poisson(shot) noise/detector noise at large $\ell$'s.
      Therefore, the measurement of the thermal SZ signal, which becomes important at large $\ell$'s,
 %     as far as the galaxy survey is concerned,
      is limited by the number of observed galaxies, as well as the resolution of the CMB detector
      (the angle at which signal-to-noise ratio for the CMB measurement is of order unity). On the other hand,
      for the small $\ell$ portion of the cross-correlation which is
      relevant for the ISW signal, the error is set by the matter
      and CMB power spectra and thus, is only limited by cosmic variance. The only way to reduce this error
      is by observing a larger volume of the universe in the redshift range $0 < z < 1$,
      where dark energy dominates.

\subsection{A Note On the Covariance Matrices}
      We saw in Section III.C that the errors in cross-correlations could be expressed in terms
      of the theoretical auto-correlation. However, this is not the whole story.

        We have a remarkable understanding of the
        auto-power spectrum of the CMB. However, if one tries to use the
        frequency information to, say subtract out the microwave
        sources, the simple temperature auto-power does not give
        the cross-frequency terms in the covariance matrix. In
        fact, in the absence of a good model, the only way to
        constrain these terms is by using the cross-correlation of
        the bands themselves. Of course, this method is limited by
        cosmic variance and hence does not give an accurate result
        at low multipoles. To solve this problem, we use the WMAP concordance model CMB auto-power
        for $\ell \leq 13$. Since there is no frequency-dependent signal at low $\ell$'s, we only use the
        W-band information, which has the lowest Galactic
        contamination \citep{mapfor}, for our first 4 $\ell$-bins which
        cover $3 \leq\ell\leq 13$ (see the end of Sec. III.D for more on our $\ell$-space
        binning).

         There is a similar situation for the contaminants of the 2MASS
         galaxy catalog. Systematic errors in galaxy counts, due
         to stellar contamination or variable Galactic extinction,
         as well as observational calibration errors,
         may introduce additional anisotropies in the galaxy
         distribution which are not easy to model. Again, the
         easiest way to include these systematics in the error is
         by using the auto-correlation of the observed galaxy distribution,
         which is inaccurate for low multipoles, due to cosmic variance. Unfortunately, this
         is also where we expect to see possible Galactic contamination
         or observational systematics. With this in mind, we try
         to avoid this problem by excluding the quadrupole, $C(2)$,
         from our analysis.

    At this point, we should point out a misconception about the nature of
    Monte-Carlo vs. jack-knife error estimates in some
    previous cross-correlation analyses, specifically
    \citep{fosalba1,fosalba2}. Many authors have used Gaussian Monte-Carlo realizations of the CMB sky
    to estimate the covariance matrix of their real-space cross-correlation
    functions\citep{boughn/crittenden:2003,fosalba1,fosalba2,scranton}.
    The justification for this method is that, since the first term
    in Eq.(23) is much smaller than the second term, the error in
    cross-correlation for any random realization of the maps is
    almost the same as the true error, and the covariance of the
    cross-correlation, obtained from many random Gaussian
    realizations is an excellent estimator of the covariance
    matrix. We may also obtain error estimates based on random realizations
    of one of the maps, as long as the observed auto-power is a
    good approximation of the true auto-power, i.e. the cosmic
    variance is low, which should be the case for angles smaller than
    20 degrees ($\ell>10$). Of course, at larger angles,
    as we mentioned above, one is eventually limited by the
systematics of the galaxy survey
    and, unless they are understood well enough, since theoretical
    error estimate is not possible, there will be no better alternative rather than Monte-Carlo error estimates.
     In fact, contrary to \citep{fosalba1,fosalba2}, if anything, the presence of
    cross-correlation makes Monte-Carlo errors a slight
    underestimate (see Eq.23). On the other hand, there is no rigorous
    justification for the validity of jack-knife covariance
    matrices, and the fact that jack-knife errors could be smaller than the
    Monte-Carlo errors by up to a factor of three \citep{fosalba1,scranton} implies that they
    underestimate the error.

    As we argue below (see Sec. III.D), since we do our analyses
    in harmonic space and use most of the sky, our $\ell$-bins
    are nearly independent and performing
    Monte-Carlo's is not necessary. Our covariance matrix is
    nearly diagonal in $\ell$-space and its elements can be
    obtained analytically, using Eq.(23).

\subsection{Angular Cross-Power Estimator}

         The WMAP temperature maps are set in HEALPix format
         \citep{healpix}, which is an equal area, iso-latitude pixellization of the
         sky. As a component of the HEALPix package, the FFT based
         subroutine `map2alm' computes the harmonic transform of
         any function on the whole sky. However, as we describe in the next section,
         in order to avoid contamination by Galactic foreground emission in
         WMAP temperature maps, and contamination by stars and
         Galactic extinction in the 2MASS survey, we have to mask
         out $\sim 15\%$ of the CMB and $\sim 30\%$ of the 2MASS
         sky. Therefore, we cannot obtain the exact values of
         the multipoles, $C_{\ell}$, and should use an estimator.

           We use a quadratic estimator which is based on the
         assumption that our masks, $W({\bf \hat{n}})$, are independent of the data
         that we try to extract (see \citep{efstathiou} for a review of different estimators). The real-space cross-correlation of the
         masked fields $\bar{A}({\bf \hat{n}})=W_A({\bf \hat{n}}) \tilde{A}({\bf \hat{n}})$
         and $\bar{B}({\bf \hat{n}})=W_B({\bf \hat{n}}) \tilde{B}({\bf \hat{n}})$ on the sphere is given by
         \bea
         &\langle\bar{A}({\bf \hat{n}})\bar{B}({\bf \hat{m}})\rangle          \langle \tilde{A}({\bf \hat{n}})\tilde{B}({\bf \hat{m}}) W_A({\bf \hat{n}})W_B({\bf \hat{m}})
         \rangle &\nonumber\\&=\langle \tilde{A}({\bf \hat{n}})\tilde{B}({\bf \hat{m}}) \rangle \langle W_A({\bf \hat{n}})W_B({\bf \hat{m}})
         \rangle,&
         \eea
          where, in the last step, we used the independence of
          data and masks, and averaged over all pairs of the same separation.
          Assuming that $\langle W_A({\bf \hat{n}})W_B({\bf \hat{m}})
         \rangle$ does not vanish for any separation (which will be true if the masked out area
         is not very large), we can invert this equation
         and take the Legendre transform to obtain the un-masked
         multipoles
         \bea
         C_{\tilde{A}\tilde{B}}(\ell) &=& \sum^{\ell_{max}}_{\ell=0} F_{\ell \ell^{\prime}}
         C_{\bar{A}\bar{B}}(\ell^\prime),~{\rm where}\nonumber\\
         F_{\ell \ell^\prime}&=&(\ell^\prime+\frac{1}{2})\int
         \frac{P_{\ell}(\cos \theta) P_{\ell^{\prime}}(\cos \theta)}{\langle
         W_A W_B \rangle(\theta)}~ d\cos\theta.
         \eea

            In fact this estimator is mathematically identical \footnote{We thank Eiichiro Komatsu for
            pointing out this identity.} to
         the one used by the WMAP team \citep{hivon}, and, within the
         computational error, should give the same result. The
         difference is that we do the inversion in real-space,
         where it is diagonal, and then transform to harmonic space, while they do
         the inversion directly in harmonic space. Indeed, using our method, we
         reproduce the WMAP binned multipoles \citep{hinshaw} within 5\%.
         However, we believe our method is computationally more
         transparent and hence more reliable.  Also, the matrix
         inversion in harmonic space is unstable for a small or
         irregular sky coverage (although it is not relevant for
         our analyses).

           Finally, we comment on the correlation among different multipoles in $\ell$-space.
         Masking about 30\% of the sky causes about 30\% correlation among
         neighboring multipoles. We bin our
         multipoles into 13 bins that are logarithmically spaced in
         $\ell$ (covering $3<\ell<1000$) , while excluding the quadrupole due to its large Galactic
         contamination in both data sets.
         The highest correlation between neighboring bins is
         $15\%$ between the first and the second bins ($C(3)$ and $(9 C(4)+11 C(5))/20$). To simplify our calculations, we neglect this correlation,
         as any correction to our results will be significantly smaller than the cosmic variance
         uncertainty(see V.B), i.e. we approximate our covariance matrix
         as diagonal in $\ell$-space.

\section{Data}

\subsection{WMAP foreground cleaned temperature maps}

  We use the first year of the observed CMB sky by WMAP for our analysis
  \citep{BennettEtAl2003a}. The WMAP experiment observes the microwave
  sky in 5 frequency bands ranging from 23 to 94 GHz. The detector
  resolution increases monotonically from 0.88 degree for the lowest
  frequency band to 0.22 degree for the highest frequency. Due
  to their low resolution and large Galactic contamination, the two bands with the lowest
  frequencies, K(23 GHz) and Ka(33 GHz), are mainly used for Galactic foreground
  subtraction and Galactic mask construction\citep{mapfor},
  while the three higher frequency bands, which have the highest
  resolution and lowest foreground contamination, Q(41 GHz), V(61 GHz), and
  W(94 GHz), are used for CMB anisotropy spectrum analysis.
   \citep{mapfor} use the Maximum Entropy
  Method to combine the frequency dependence of 5 WMAP bands with
  the known distribution of different Galactic components that
  trace the dominant foregrounds to obtain the foreground
  contamination in each band. This foreground map is then used to
  clean the Q, V and W bands for the angular power spectrum
  analysis. Similarly, we use the cleaned temperature maps of
  these three bands for our cross-correlation analysis. We also
  use the same sky mask that they use, the Kp2 mask which masks out 15\% of the sky,
  in order to avoid any remaining Galactic foreground, as well as 208 identified
  microwave point sources.

\subsection{2MASS extended source catalog}
We use galaxies from the Near-IR \emph{Two Micron All Sky Survey}
\citep[2MASS;][]{Skr97} as the large-scale structure tracer of the
recent universe. Our primary data set is the public full-sky
extended source catalog \citep[XSC;][]{Jar00}. The $K_s$-band
isophotal magnitude, $K_{20}$, is the default flux indicator we
use to select the external galaxies for our analysis. $K_{20}$ is
the measured flux inside a circular isophote with surface
brightness of 20 mag ${\rm arcsec^{-2}}$. The raw magnitudes from
the catalog were corrected for Galactic extinction using the IR
reddening map of \citet{Sch98}: \beq K_{20} \rightarrow K_{20} -
A_K, \eeq where $A_K = R_K E(B-V) = 0.367 \times E(B-V)$
\footnote{This is different from $R_K = 0.35$ used by
\citep{Koc01} whose luminosity function parameters we use to
estimate the redshift distribution, but the median difference of
extinction derived between the two is small($< 0.002$ mag).}.
There are approximately 1.5 million extended sources with
corrected $K_{20} < 14.3$ after removing known artifacts
(cc\_flag != 'a' and 'z') and using only sources from a uniform
detection threshold (use\_src = 1).

\subsubsection{Completeness and Contamination}
\begin{figure}[t]
\includegraphics[width=300pt]{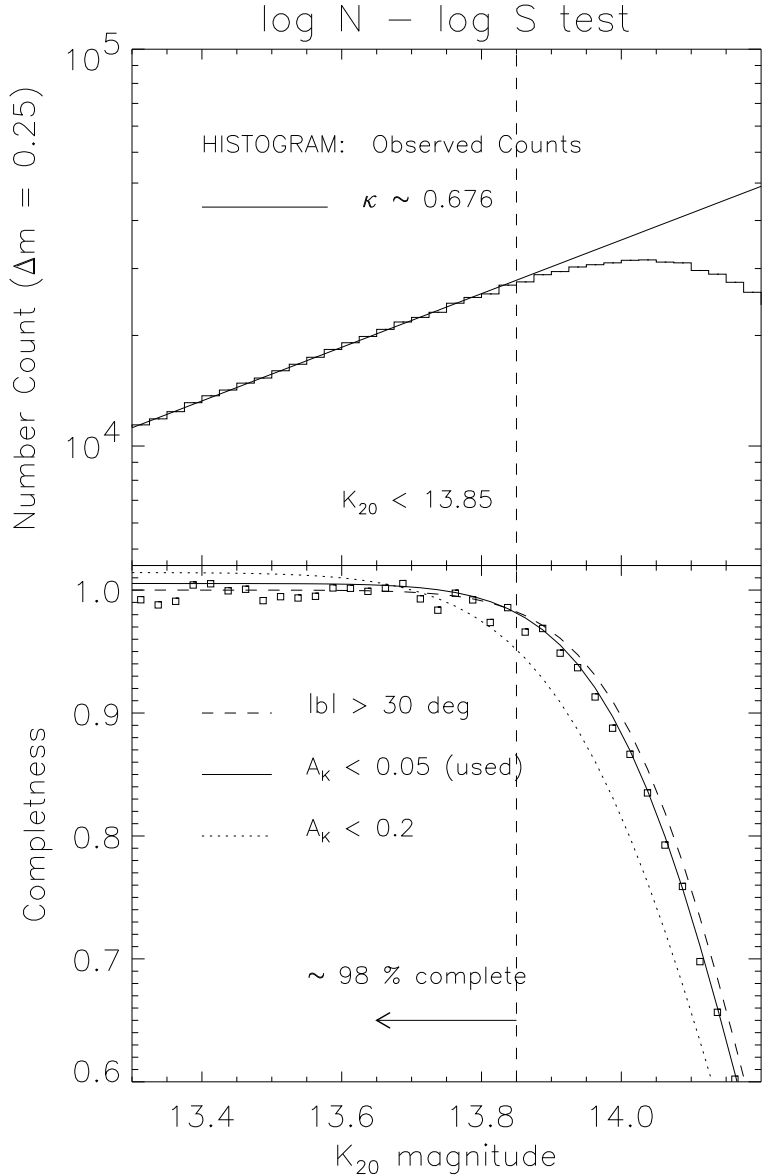}
\caption{\label{fig:num_counts} (Top panel) The histogram is the
observed $K_{20}$ number-magnitude relation for galaxies in
regions with $A_K < 0.05$. The solid line is the model counts
inferred using data from $|b| > 30^{\circ}$ in the magnitude range
$13.2 < K_{20} < 13.7$ where the extended source catalog(XSC) is
most reliable. (Bottom panel) The square points gives the
completeness as inferred from the difference between the
observed and model counts. The \emph{solid} curve is a fit to a
parametric model that estimates both the incompleteness and
contamination rate in a consistent manner. The \emph{dotted} curve
is a similar fit using data with a less stringent $A_K < 0.2$. The
\emph{dashed} curve is from $|b| > 30^{\circ}$, which serves roughly as
the completeness upper-bound for the XSC. The vertical line at
$K_{20} = 13.85$ gives a completeness at $98 \%$ for data with
$A_K < 0.05$ used in our analysis.}
\end{figure}

We use the standard $\log$ N-$\log$ S test to determine the
completeness limit of the extended source catalog. The top panel
of Figure \ref{fig:num_counts} shows the number of galaxies as a
function of $K_{20}$. The $\log$ number counts can be approximated by a
power-law: \beq \frac{dN}{dm} \propto 10^{\,\kappa\,m}.
\label{eq:numcounts} \eeq To infer the true number count-magnitude
relation, we need to ensure that our catalog is free from
contaminants since not all extended sources from the XSC are
external galaxies. At low Galactic latitude where stellar density
is high, unresolved multiple star systems are often confused as
extended sources.
%The efficiency of source detection is also considerably
%lower since foreground would scatter an otherwise robust sources
%away from the detection threshold.
For the purpose of fitting for the power-law slope $\kappa$, we
use only sources with $|b| > 30^{\circ}$. Using $\sim 250,000$
galaxies in the magnitude range $13.2 < m < 13.7$ (where the
reliability has been determined to be $99 \%$ by \citet{Huc01}),
we fitted a number count slope $\kappa = 0.676 \pm 0.005$.

While the XSC is unreliable at low Galactic
latitudes, the $|b| > 30^{\circ}$ cut is too
permissive and would throw away a large area of the sky that can
be used for analysis. In principle, we could use the stellar density
$n_{star}$ from the 2MASS Point-Source Catalog(PSC) to set a
threshold for excluding region of high stellar density. However,
since it has been shown by \citep{Bel03} and \citep{Ive00} that
unresolved extended galaxies are found in the PSC(up to $2 \%$ of
all point sources with $K_{20}\sim 14$), a mask derived from the
observed stellar density would preferentially exclude regions of
high galaxy density.

\begin{figure}[t]
\includegraphics[angle=90, width=450pt]{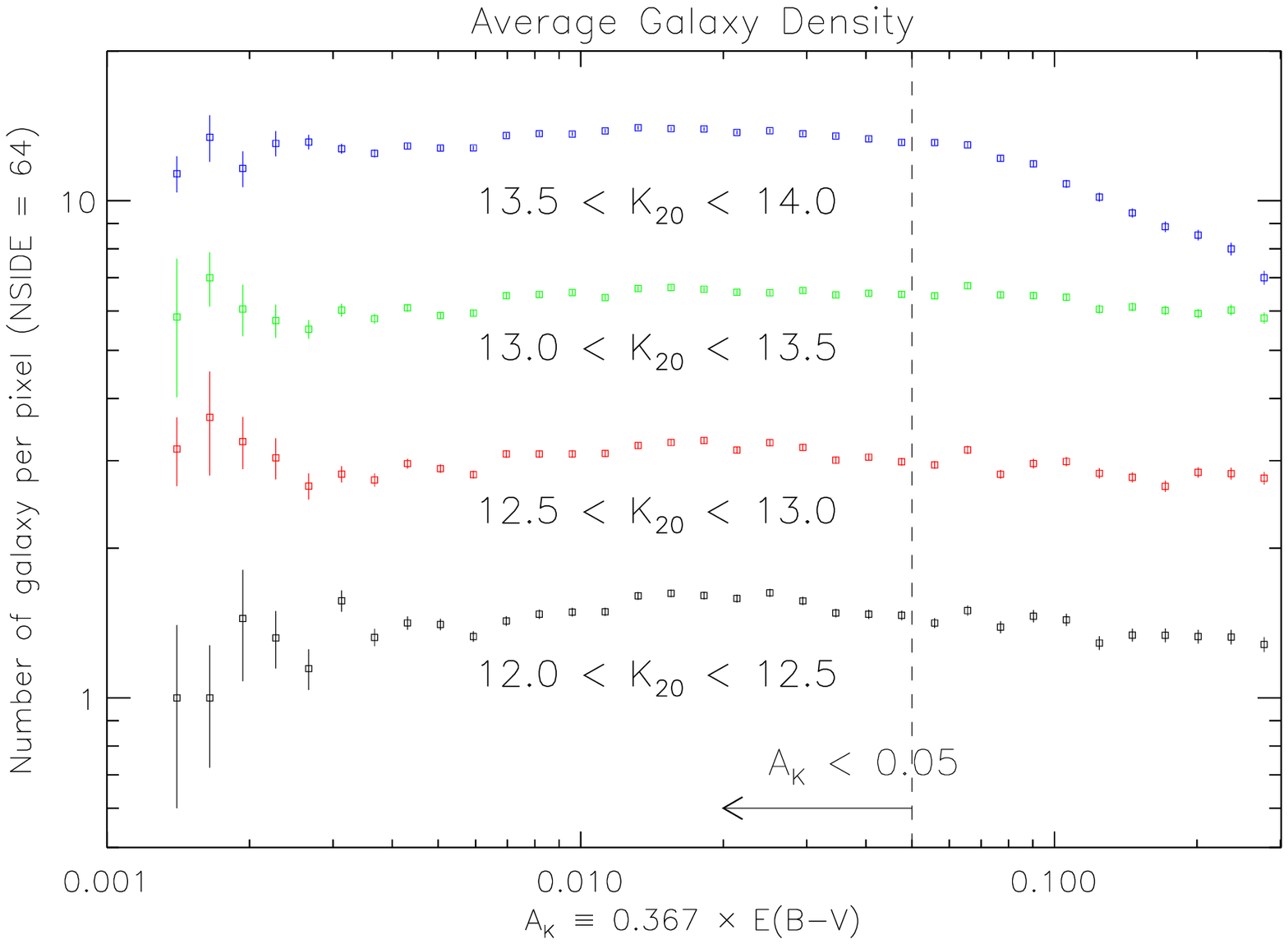}
\caption{\label{fig:nbar_ak} Average number of galaxies per
$0.83\,\,{\rm deg}^2$ pixel (HEALpix $N_{side}$ = 64) as a
function of extinction. For bright galaxies ($K_{20} < 13.5$), the
galaxy density is constant up to extinction value $\sim 0.25$. For
$13.5 < K_{20} < 14.0$, the density drops off at $A_K \sim 0.65$.
We use only regions with $A_K < 0.05$ (\emph{dashed vertical
line}) for our analysis. Errors are estimated using jack-knife
resampling.}
\end{figure}

We use the extinction map of \citep{Sch98} to exclude regions of
the sky where the XSC is unreliable. Figure \ref{fig:nbar_ak}
shows the average number of galaxies per HEALPix pixel of
$0.83\,{\rm deg}^2$ ($N_{side} = 64$), as a function of Galactic
extinction for the four magnitude ranges used in our analysis. For
bright galaxies, e.g. $K_{20} < 13.5$, the Galactic density is
constant on degree scales. For the faintest magnitude bin, the
number density drops off at large $A_K$ for $A_K$ beyond $\sim
0.065$. We thus choose $A_K < 0.05$ \footnote{This is also the
level chosen by \citep{Mal03} for their auto-correlation analysis
of the XSC.}. This stringent threshold excludes $\sim 99\%$ of all
regions with $n_{star} > 5000\,{\rm deg}^{-2}$. Moreover, it
improves the global reliability of galaxy counts, as our flux
indicator $K_{20}$ for each source was corrected for Galactic
extinction, which has an uncertainty that scales with $A_K$
itself. This cut reduces the number of extended sources with
$K_{20} < 14.3$ to $\sim 1$ million, covering $\sim 68.7 \%$ of
the sky. For the sake of completeness, we also repeat our
cross-correlation analysis for a less stringent mask with $A_K <
0.1$, which covers $\sim 79.0 \%$ of the sky.

Using $\kappa = 0.676$ derived from regions with $|b| >
30^{\circ}$ as a model for the true underlying number counts, we
infer the catalog completeness and contamination as a function of
apparent magnitude for the extinction cropped sky. We deduce the
intercept of the linear $\log$ counts - magnitude model by scaling
the observed number counts from the $|b| > 30$ region to the
larger $A_K$ masked sky at the bright magnitude range $12.5 <
K_{20} < 13.0$. Essentially, we assumed the two number count
distributions are identical at those magnitudes. The observed
fractional deviation from Eq. (\ref{eq:numcounts}) \beq I(m)~ =
~\left({\frac{dN}{dm}}^{\kappa} \,-\, {\frac{dN}{dm}}^{\rm
obs}\right)\Big/
               {\frac{dN}{dm}}^{\kappa}\label{eq:dev}
\eeq
is positive at faint magnitudes indicating incompleteness but
crosses zero to a constant negative level towards the bright-end,
which we inferred as contamination to the XSC. Plotted in the bottom
panel of Figure \ref{fig:num_counts} is the completeness function
$C(m) \equiv 1 - I(m)$, where we parametrically fitted using
\beq I(m) ~=~ I_{o}\exp\left[-\frac{(m - \bar{m})^2}{2\sigma^2}\right]
              - {\it Const}. \label{eq:dev2}
\eeq
In Figure \ref{fig:num_counts}, the term \emph{Const} describes the low
level of excursion beyond $C(m) = 1$.
We obtained a $\sim 98 \%$ completeness for $K_{20} < 13.85$ and
contamination rate at $0.5 \%$ level for $A_K < 0.05$ (solid curve).
As a comparison, a less stringent threshold of $A_K < 0.2$ (dotted
curve), the completeness is $\sim 95 \%$ with contamination at
$1.5 \%$. The dashed curve is computed using high latitude data
($|b| > 30^{\circ}$), serves roughly as the completeness
upper-bound (as a function of apparent magnitude) for the XSC.

At a low level, contaminants in the catalog merely increase the
noise of our signal with marginal systematic bias. The $A_K <
0.05$ extinction mask is close to optimal in terms of
signal-to-noise for our cross-correlation analysis. One the other
hand, catalog incompleteness at faint magnitudes affects our
ability to infer the correct redshift distribution. We use
galaxies up to $K_{20} = 14.0$ but weighted the redshift
distribution at a given magnitude range (described below) by Eq.
(\ref{eq:dev2}).

\subsubsection{Redshift Distribution}
The redshift distribution of our sample was inferred from the
\citet{Sch76} parameters fit of the $K_{20}$ luminosity function
from \citep{Koc01}. The redshift distribution, $dN/dz$ of galaxies
in the  magnitude range $m_{\bright} < m < m_{\faint}$ is given by
the integration of the luminosity function $\Phi(M)$
\beq
\frac{dN}{dz}(z)\,dz =
\int_{M_b(z)}^{M_f(z)}\,\Phi(M)\,dM
             \times\frac{dV_c}{dz}(z)\,dz, \label{eq:dn_dz}
\eeq where $dV_c/dz$ is the line-of-sight comoving volume element
and \bea
M_f(z) &\equiv& m_{\faint} - DM(z) - k(z)\\
M_b(z) &\equiv& m_{\bright} - DM(z) - k(z)
\eea
Here, $DM(z)$ and $k(z)$ are the distance modulus and
\emph{k}-correction at redshift z. To be consistent with \citep{Koc01},
we employ $k(z) = -6.0 \log (1 +z)$, but the redshift distribution is
insensitive to the exact form of the \emph{k}-correction.
%using other forms of \emph{k}-correction like $k(z)\sim
%-2.1\pm0.3z$\citep{Bel03} or $k(z)\sim -2.25z$\citep{Gla95}
%changes both the peak and the shape of the derived redshift
%distribution only slightly.
The Schechter parameters used were
%$\phi^* = 1.16 \times 10^{-2}\,\ihmpcC $,
$M^* = -23.39$ and $\alpha_{s} = -1.09$. For analytic convenience,
we further model $dN/dz$ with the three parameter
generalized-gamma distribution: \bea
\frac{dN}{dz}(z\,|\,\lambda,\beta,z_o)\,dz
&\propto&\,\frac{\beta}{\Gamma(\lambda)}
                                        \left(\frac{z}{z_o}\right)^{\beta\lambda - 1}
                                        \nonumber\\
             &\times& \exp\left[-\left(\frac{z}{z_o}\right)^{\beta}\right]\,
                                    d\left(\frac{z}{z_o}\right).
\eea The fit were weighted by relative counts, hence they are exact
near $z_o$, the mode of the distribution, but underestimate the
true number at the high redshift tail by less than $1 \%$.
%begin table 1
\begin{table}
\begin{center}
\caption{\label{tab:dn_dz_parm}
 $\,\frac{dN}{dz}$ parameters for the four magnitude bins}
\begin{tabular}{c c c c c}
\hline \hline
                        & $z_o$ & $\beta$ & $\lambda$ & $N$ \\
\hline
$12.0 < K_{20} < 12.5\quad\quad$   & 0.043~ & 1.825~ & 1.574~ & 49817 \\
$12.5 < K_{20} < 13.0\quad\quad$   & 0.054~ & 1.800~ & 1.600~ & 102188 \\
$13.0 < K_{20} < 13.5\quad\quad$   & 0.067~ & 1.765~ & 1.636~ & 211574 \\
$13.5 < K_{20} < 14.0\quad\quad$   & 0.084~ & 1.723~ & 1.684~ & 435232 \\
\hline
\end{tabular}
\end{center}
\end{table}
%end of table 1
Table \ref{tab:dn_dz_parm} gives the redshift distribution
parameters for the four magnitude bins used in our analysis. We
normalize the integral of $dN/dz$ with the total number of
observed galaxies in the respective magnitude range, \bea
N_{\total}(\Delta \Omega)&=&\int_0^{\infty}\,\frac{dN}{dz}(z)\,dz\times \Delta \Omega \\
          &=&\int_0^{\infty}\,n_c(z)\,\frac{dV_c}{dz}(z)\,\,dz\int d\Omega\nonumber\\
          &\equiv&\int n_c(r)~r^2\,dr\,d\Omega. \label{eq:metric}
\eea The observed $N_{\total}$ is consistent with the 10\%
uncertainty in the normalization, $\phi^{*} = 1.16 \times
10^{-2}\,\ihmpcC $ obtained by \citep{Koc01} in their luminosity
function analysis. Eq. (\ref{eq:metric}) gives the explicit
relation (in the absence of clustering) between the redshift
distribution, $dN/dz$, and the comoving density, $n_c(r)$ used in
Sec. III.A.

\begin{figure}[t]
\includegraphics[angle=90,width=450pt]{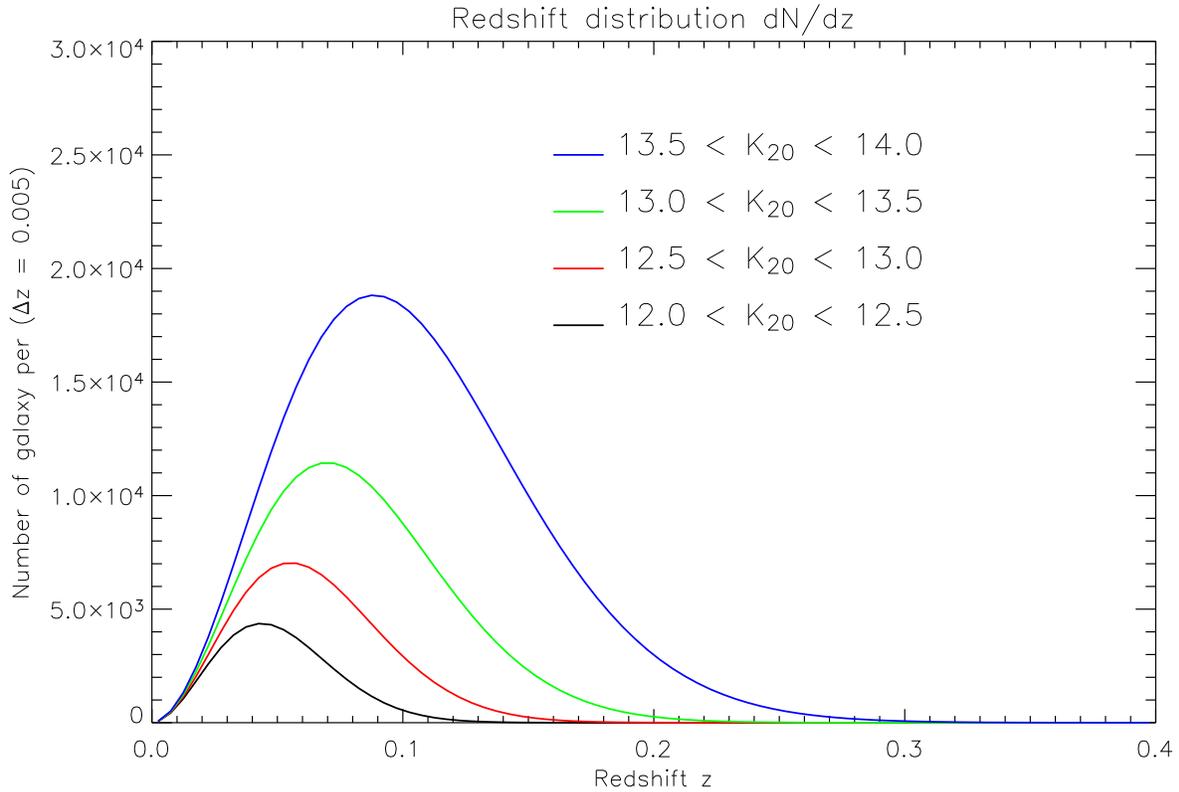}
\caption{\label{fig:dn_dz} $dN/dz$ for the four magnitude bins
used in the analysis.}
\end{figure}

Figure \ref{fig:dn_dz} is a plot of the redshift distribution for
the four magnitude bins used in our analysis. For the first three
bright samples, where we are complete, the parameters for $dN/dz$
were derived from a direct application of Eq. \ref{eq:dn_dz}. For
$13.5 < K_{20} < 14.0$, the redshift distribution was computed by
summing up magnitude slices with interval $\Delta K_{20}=0.05$,
and weighted by their relative number counts.

\section{Results}
For the following results, unless we mention otherwise, we use the
WMAP concordance cosmological model.

\begin{figure}[t]
\includegraphics[width=450pt]{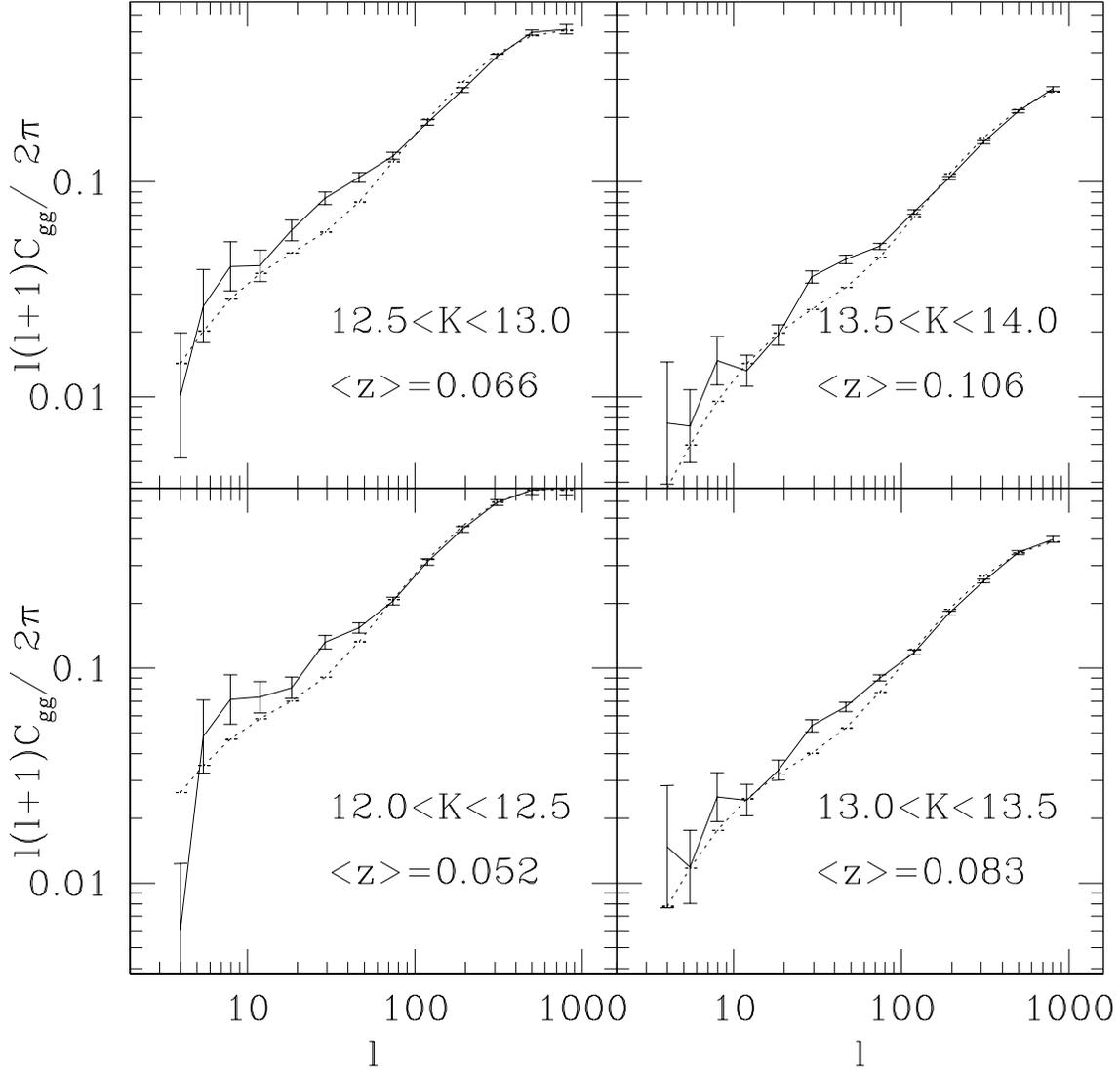}
\caption{The auto-power for our four different magnitude bins. The
solid curves show the observed auto-power multipoles with their
estimated Gaussian errors (Eq.23), while the dashed curves are the
projected Peacock and Dodds \cite{peacock} non-linear power
spectra with the best fit constant bias. The best fit Poisson
noise term is subtracted out.}
\end{figure}

By comparing the angular auto-power spectrum of the galaxies in
each magnitude bin with the theoretical auto-power spectrum
(Eq.19), we can obtain the bias of the 2MASS galaxies. In order to
do this, we use a $\chi^2$ fit, assuming independent gaussian
random errors at each $\ell$-bin. Figure 4 compares our best fit
models of the auto-power (solid curves) with the measured
auto-powers for each magnitude bin. The value of the bias for all
the magnitude bins is within \beq b_g =1.11 \pm 0.02,\eeq which
confirms our constant bias assumption\footnote{Given that the
galaxy distribution is non-linear and non-gaussian, the $\chi^2$
fit is not the optimal bias estimator. However, the fact that the
biases for different bins are so close implies that the error in
bias, as we see below, is much smaller than the error in our
cross-correlation signal and so is negligible.}. Our values for
the Poisson correction factor (see Eq.19), $\gamma$, are all
within 1\% of $1.02$. The most significant deviation of the
theoretical fit from the observed auto-power is about 30\% at
$\ell\sim 30-40$. One possibility may be that galaxy bias is
larger at large (linear) scales than at the (non-linear) small
scales. In order to estimate the effect, we can limit analyses to
the first 7 $\ell$-bins ($\ell \lesssim 70$, scales larger than
$\sim 7-13 ~h^{-1}\mpc$). This yields the estimated bias on linear
scales: \beq b_{g,{\rm lin}} = 1.18 \pm 0.08, \eeq
%As to the source of this deviation, the most significant
%deviations are at the scale of a few degrees which is close the
%width of the 2MASS scanning stripes(6 degrees)(ref?). The
%amplitude of deviation implies systematic fluctuations of order
%10\% in the number counts. If due the errors in the photometric
%zero-point, such fluctuations require a magnitude error $\Delta m
%\sim 0.06$, which is significantly larger than $\Delta m <0.02$
%claimed by \citep{Nikolaev2000}. Although we do not have an
%explanation for the source of this discrepancy, it is conceivable
%that a combination of observational systematics may lead to 10\%
%fluctuations at the characteristic scale of the 2MASS survey.

The angular scale corresponding to $\ell =30-40$ is a few degrees,
which is close to the length of the 2MASS scanning stripes
($6^\circ$). The amplitude of deviation from the constant bias
model would require systematic fluctuations of order 10\% in the
number counts on that scale. If these were due to systematic
errors in the 2MASS photometric zero-point, such fluctuations
would require a magnitude error $\Delta m \sim 0.06$, which is
significantly larger than the calibration uncertainties in 2MASS
\citep{Nikolaev2000}. Therefore, we will use our estimated linear
bias (Eq. 37) for the interpretation of our ISW signal, while we
use the full bias estimate (Eq. 36), which is dominated by
non-linear scales, to analyze our SZ signal.

 The points in Figure 5 summarize our twelve observed
cross-correlation functions (3 WMAP bands $\times$ 4 magnitude
bins), while Figure 6 shows the same data after subtracting out
the best fit contribution due to microwave Point Sources. We fit
our theoretical model (Eq. 18) to our cross-correlation points
(including only the W-band for the first 4 $\ell$-bins; see Sec.
III.C), allowing for free global normalizations for the ISW, SZ
and Point Source components. The curves show this model with the
best fit normalizations for these components, while the shaded
region shows the 68\% uncertainty around a null hypothesis. Figure
7 shows how individual theoretical components depend on frequency
and $\ell$ for our faintest magnitude bin.
 As we mentioned in Sec. III.C, different $\ell$-bins are nearly
 independent. However, different combinations of frequency bands
 and magnitude bins are highly correlated and we use the full
 covariance matrix which we obtain from the data itself (see
 III.C) for our $\chi^2$ analysis.

 The apparent dispersion in our data points for
  the first 4-5 $\ell$-bins is smaller than what we expect from
  gaussian statistics (the shaded regions in Figures 5 and 6).
  This may be due to the non-gaussian nature of the systematics
  (observational or Galactic), which dominate the error on large
  angles, and make the variance (Eq. 23) significantly different from the
  $68\%$ confidence region.

 Figures 5 \& 6 show that our faintest magnitude bin has
 the smallest error. This is due to the
 fact
 that our faintest magnitude bin covers the largest effective comoving volume and number of
 galaxies (see Table I and Figure 3), and as a result, between 50-70\% of our signal
 (depending on the component) comes from this bin. Repeating the
 statistical analysis for individual magnitude bins leads to results consistent
 with the combined analysis within the errors.

\begin{figure}[t]
\includegraphics[width=450pt]{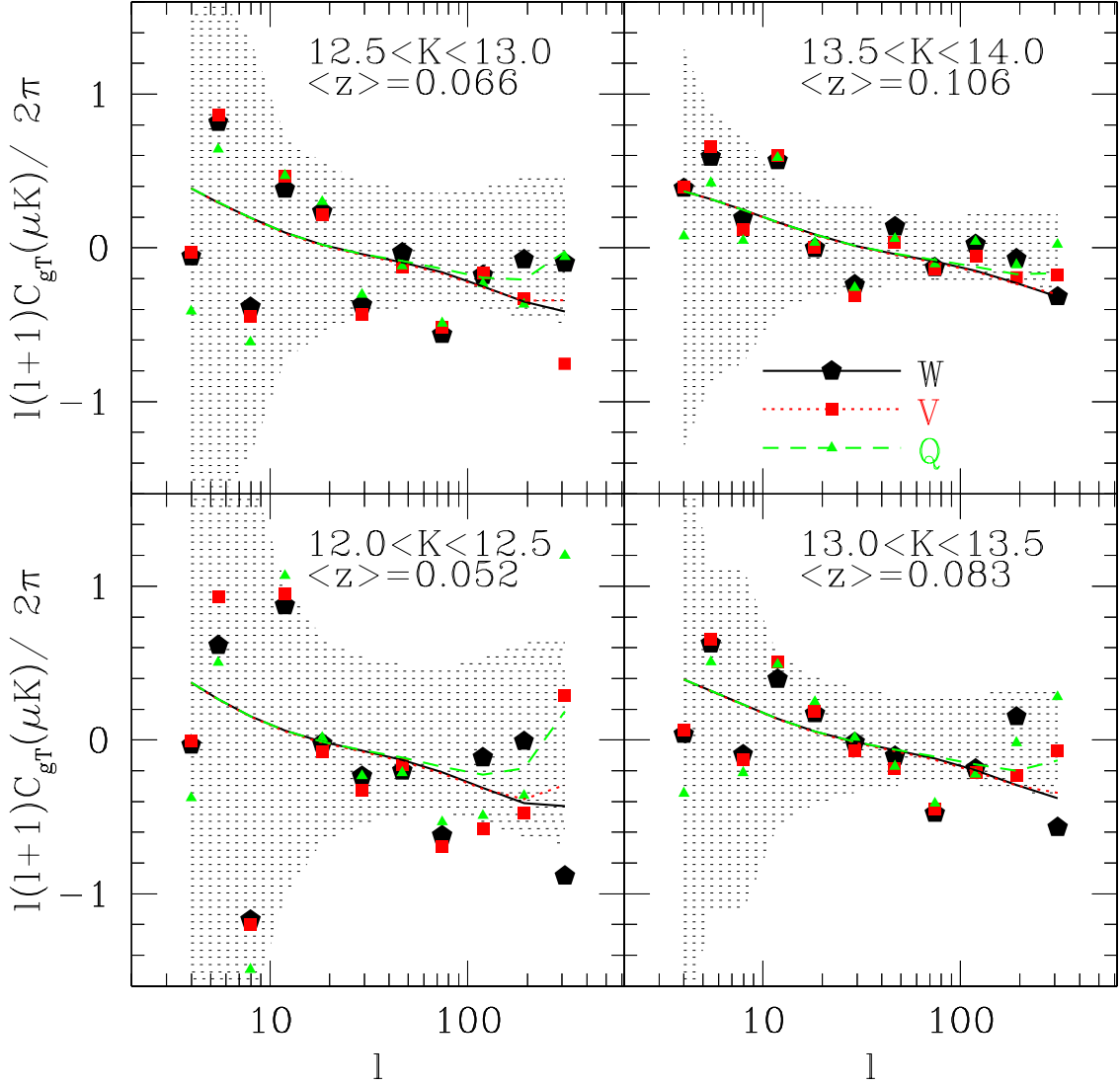}
\caption{The cross-power for our four magnitude bins. The curves
are the best fit model (ISW+SZ+Point Sources) for the three bands
and the points show the data. The ISW/SZ components dominate the
signal for $\ell$'s below/above 20. The Point Source contribution
becomes important for the lower frequency bands at the highest
$\ell$'s. The shaded region shows the $1-\sigma$ error centered at
the null hypothesis. Note that, while different $\ell$-bins are
nearly independent, different cross-powers of bands with magnitude
bins are highly correlated. As shot noise dominates the signal for
our last two $\ell$-bins, for clarity, we only show the first 11
$\ell$-bins, for which the errors for the three WMAP bands are
almost the same.}
\end{figure}

\begin{figure}[t]
\includegraphics[width=450pt]{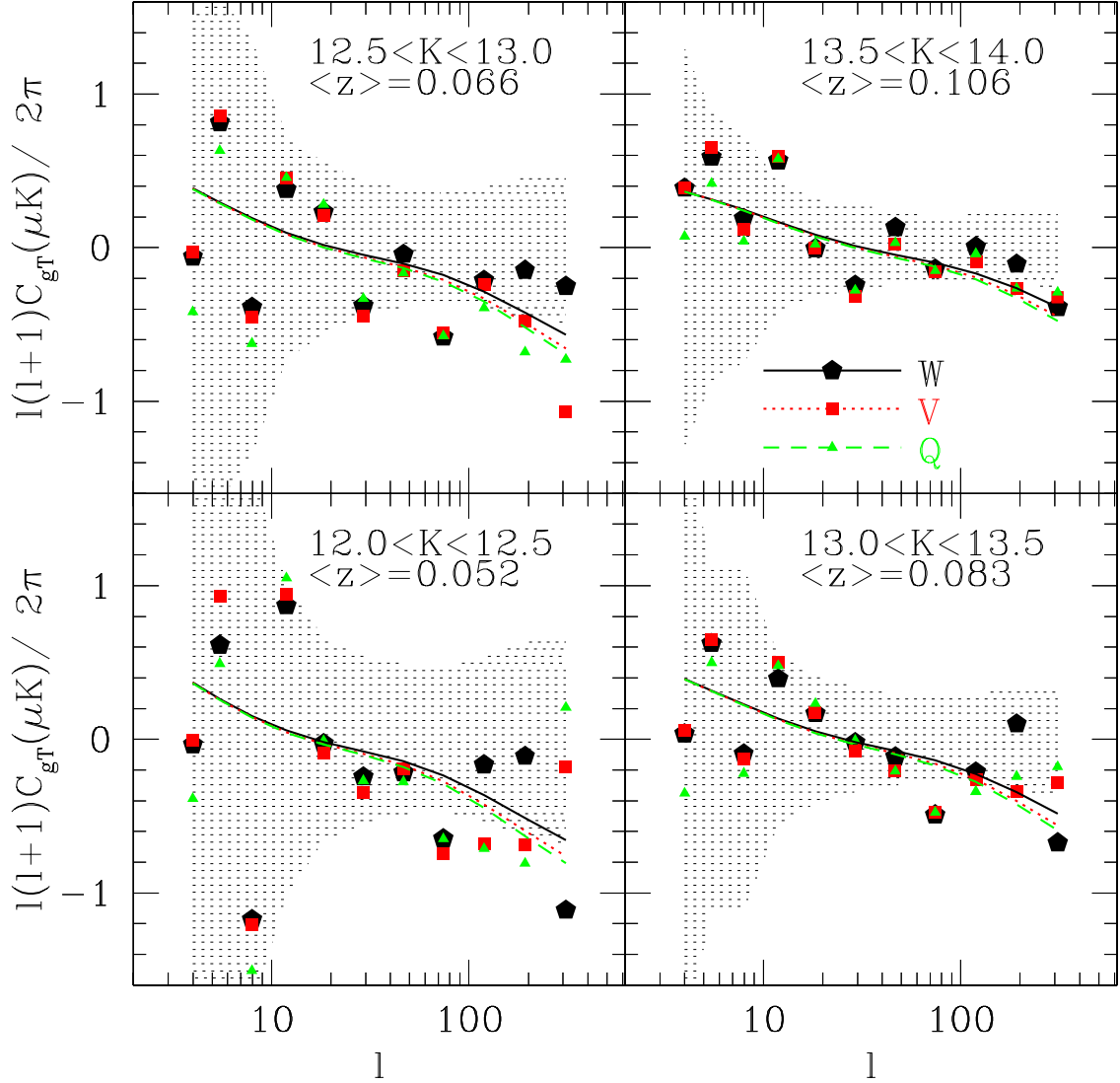}
\caption{The same as Figure 5, but with the Point Source
contributions subtracted from both theory and data.}
\end{figure}

\begin{figure}[t]
{\includegraphics[width=200pt]{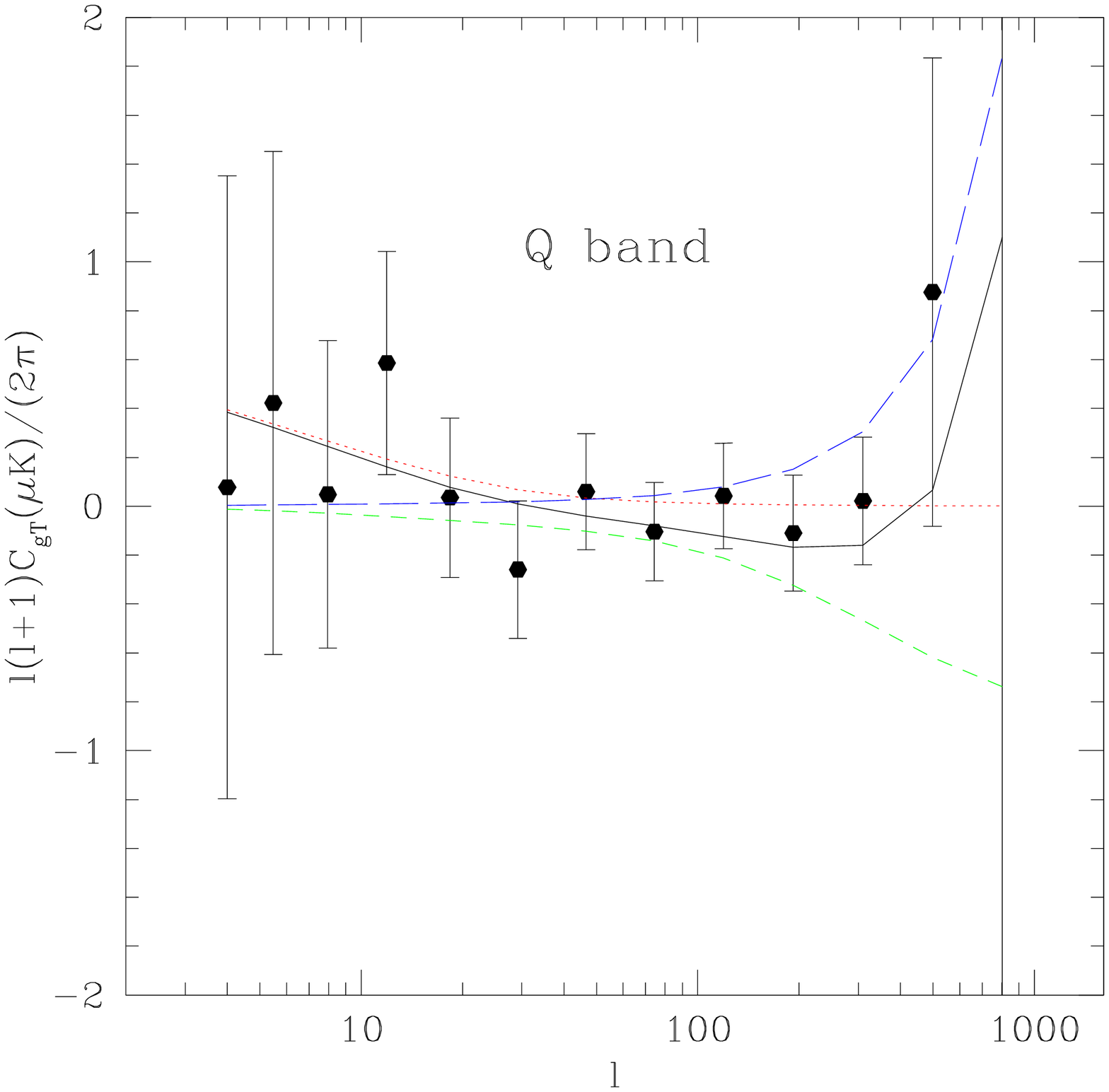}
\includegraphics[width=200pt]{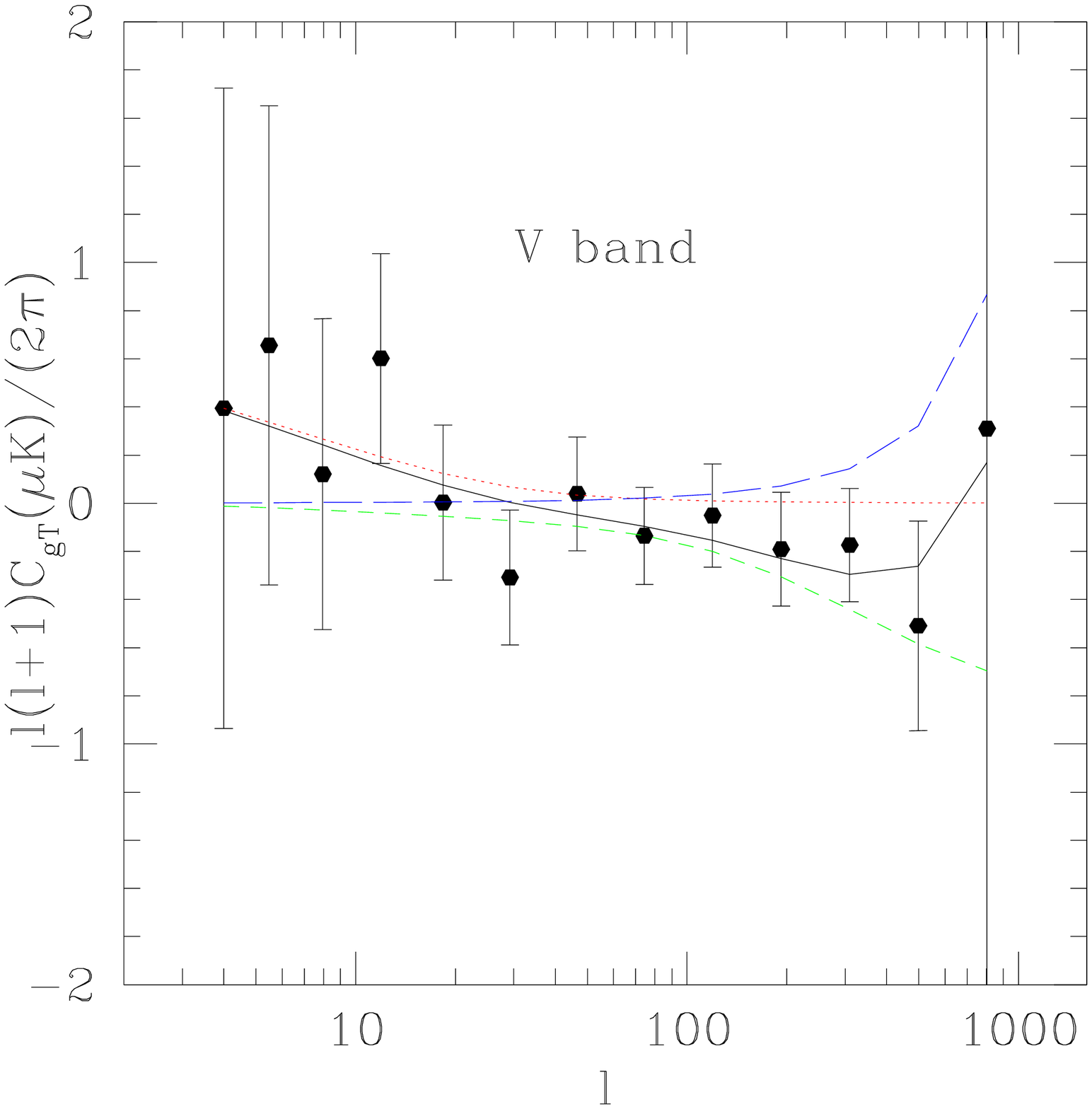}
\includegraphics[width=200pt]{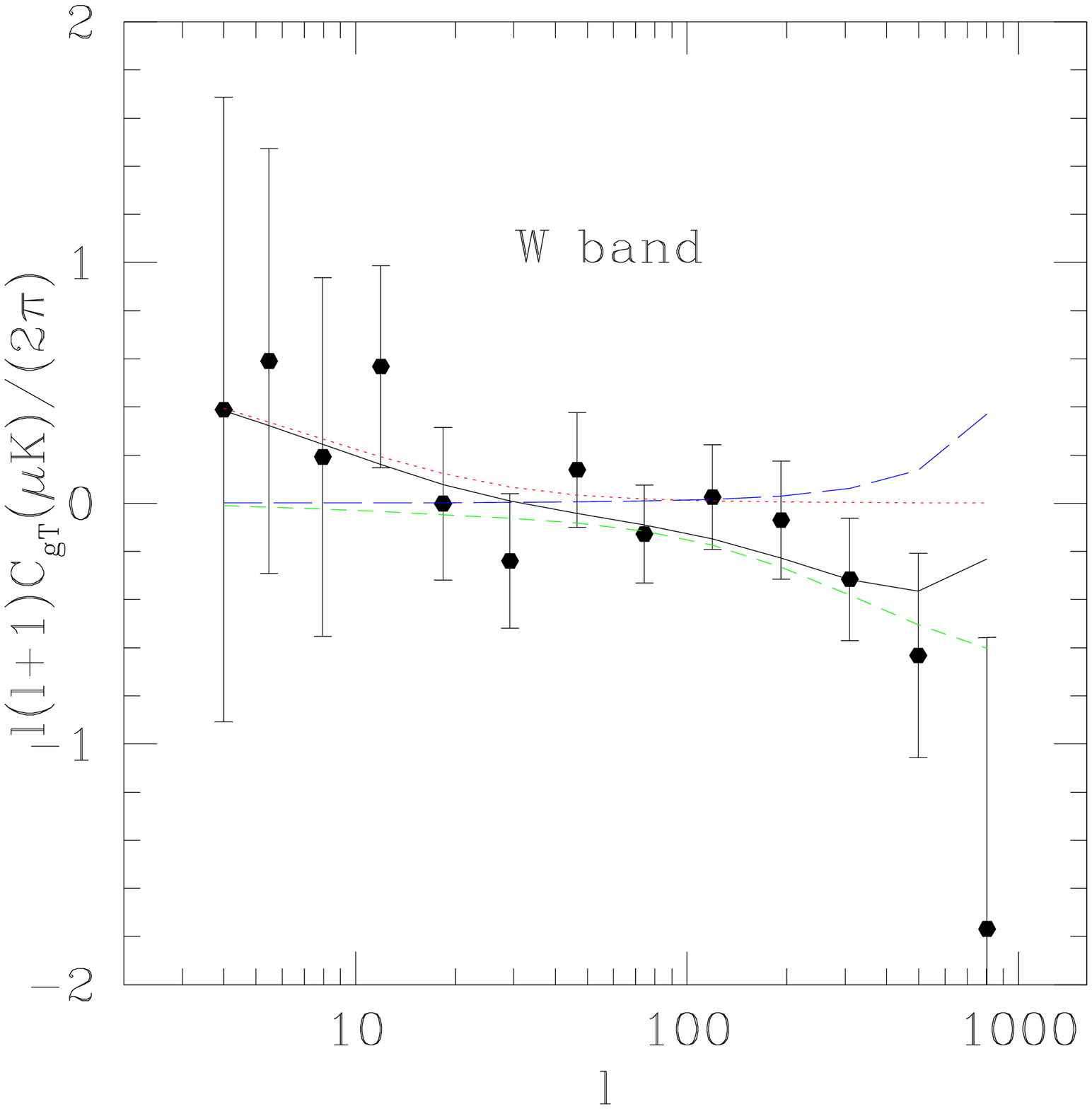}}
\caption{Different components of our best fit theoretical
cross-power model, compared with the data for our faintest
magnitude bin ($13.5<K<14$). The dotted(red) curves show the ISW
component, while the short-dashed(green) and long-dashed(blue)
curves are the SZ and Point Source components respectively. The
black curves show the sum of the theoretical components, while the
points are the observed cross-power data.}
\end{figure}

\subsection{Thermal SZ signal}
   In a model described in Appendix A, we quantify the amplitude of the
   SZ signal in terms of $Q$, the coefficient of temperature-mass
   relation for clusters of galaxies
   $$T_e(M)=(6.62 ~\kev)Q \left(\frac{M}{10^{15} h^{-1}
    M_{\odot}}\right)^{2/3}.\eqno(A2)
   $$
   Different theoretical and
   observational methods place $Q$ somewhere between $1$ and $2$,
   with observations preferring the higher end. This amplitude
   could be equivalently described in terms of $\tilde{T}_e$,
   the product of pressure bias and average electron temperature (see Eq.7), which is less model dependent.
   Our best fit value for the thermal SZ signal, which shows a signal at the $3.1\sigma$ level, is
   \beq
   Q = 1.19 \pm 0.38, ~{\rm or} ~ \tilde{T}_e =b_p \bar{T}_e  =
   (1.04
   \pm 0.33)~\kev,
   \eeq
   which is consistent with the X-ray observations of galaxy
   clusters.

   This result is slightly dependent on the spectrum of the microwave
   point sources, which we discuss below in V.C. If we restrict
   the analysis to $\ell>20$, which is where all the SZ signal comes
   from, and our estimates of the covariance matrix is robust,
   our reduced $\chi^2$ is $0.93$ which is within the
   68\% allowed range for our $12\times 13$ degrees of freedom. This
   implies that there is no observable systematic deviation from our
   theoretical expectation for the shape of the thermal SZ cross-power
   (or its gaussianity).

     Repeating the analysis with the $A_K<0.1$ extinction mask (See Sec. IV.B.1) for the 2MASS galaxies,
     which has a 10\% larger sky coverage, increases the SZ signal
     slightly to $Q=1.27 \pm 0.35$, which is a detection at the $\sim 3.7 \sigma$
     significance level. This is probably because the Galactic
     contamination close to the plane is only at large angles
     and does not contribute to the SZ signal. Therefore, as long as the
     Galactic contamination does not completely dominate the
     fluctuations, increasing the area only increases the SZ signal.

%    Our low level of detection for the thermal SZ effect is
%    consistent with the non-detection of a WMAP cross-correlation with the ROSAT X-ray map
%    in \citep{diego}.

%   If we take
%   this $2 \sigma$ deviation from the canonical value at face value, there could, at least,
%   be four possible explanations:

 \subsection{ISW signal}

  Using our estimated linear bias (Eq. 37), our $\chi^2$ fit yields an ISW signal of

  \bea
 &  {\rm ISW} = 1.49 \pm 0.61 &\\
 &~\times {\rm ~ concordance~ model ~prediction},&\nonumber
  \eea
  a $2.5\sigma$ detection of a cross-correlation.  As with the
  previous cross-correlation analyses\citep{Bou02,fosalba1,fosalba2,Nol03}, this is consistent
  with the predictions of the concordance \lcdm paradigm. However, among the three signals
  that we try to constrain, the ISW signal is the most
   difficult to extract, because almost all the signal
   comes from $\ell<20$, given our redshift distribution. For such low
   multipoles, there are several potential difficulties:

  1- The small Galactic contamination or observational systematics
  in 2MASS may dominate the
fluctuations in the projected galaxy density at low multipoles and
wipe out the signal. However, since we use the observed auto-power
of 2MASS galaxies for our error estimates, this effect, which does
contribute to the auto-power (and not to the signal), is included
in our error.

  2- Our covariance estimator loses its accuracy as the cosmic
  variance becomes important at low multipoles (see III.C). A random error in the covariance matrix
  can systematically increase the $\chi^2$ and hence decrease the estimated error of our signal.
  However, our reduced $\chi^2$ is $0.88$, which is in fact on the low
  side (although within $1\sigma$) of the expected values for 124
  degrees of freedom \footnote{A low $\chi^2$ is to be expected if
  we overestimate the noise in the ISW signal. In particular, this could be the case if
  the CMB power is suppressed on large angles. This would increase the significance of
  an ISW detection at such scales. This effect is elaborated in \citep{kamion}}
  (remember that we only used the W band for
  the first 4 $\ell$-bins). Assuming gaussian statistics, this
  implies that we do not significantly underestimate our
  error.

  3- Possible Galactic contamination in WMAP may correlate with
  Galactic contamination in 2MASS at low multipoles, which may lead to a fake
  positive signal. However, the largest contribution of Galactic foreground is
  visible in the Q-band \citep{mapfor}, and our low $\ell$ multipoles have in fact a lower amplitude in Q band.
  Although this probably shows a large error due to contamination in the
  Q-band amplitude, the fact that this is lower than the amplitude of V and
  W bands implies that our main signal is not contaminated.
  Because of the reasons mentioned in III.C, we only use the
  W-band information for $\ell <14$.

    Using the less stringent extinction mask, $A_K<0.1$ (see IV.B.1), for the 2MASS sources yields
    a signal of ISW$=1.9 \pm 1.1$, which is a lower signal to noise detection at the
    $1.7\sigma$ level. This is probably due to the fact that most of the ISW signal comes from
    angles larger than $\sim 10^\circ$, which is highly
    contaminated in regions close to the Galactic Plane.

    Finally, we should mention that since the ISW signal comes
    from small $\ell$'s, while the SZ and point source signals come
    from large $\ell$'s (See Figure 7), there is a small correlation (less than 10\%)
    between the ISW and other signals.
 \subsection{Microwave Point Sources}

  \begin{table}[t]
\begin{center}
\caption{Best fit point source strengths for different assumed
spectra. The associated best fit SZ signal and $\chi^2$ are also
quoted. Here, $T_A$ stands for the antenna temperature, while
$L^*_V$, defined in Eq.(B2), is the estimated luminosity of the
Milky Way in WMAP's V-band.}
   \begin{tabular}{@{\extracolsep{\fill}}c||c|c|c}
   Spectrum & $L_V/L^*_V$ & $Q$ & $\chi^2$\\
  \hline
  Milky Way & $16.2 \pm 7.8$ & $1.10 \pm 0.40$ & $111.2$ \\
  $\delta T_A \propto \nu^{-2}$ & $21.0 \pm 8.1$ & $1.19 \pm 0.38$ &
  $109.5$ \\
  $\delta T_A \propto \nu^{-3}$ & $10.9 \pm 4.7$ & $0.94 \pm 0.33$ &
  $110.8$
   \end{tabular}
\end{center}
\end{table}
    As described in II.C, we
   assume that our point sources trace the 2MASS objects and have
   either a Milky Way spectrum, a $\nu^{-2}$, or a $\nu^{-3}$
   frequency dependence their antenna temperature (the last two are the expected synchrotron spectrum
   of radio sources).
   The results are shown in Table II. We see that, although all the spectra are consistent
   at a $2\sigma$ level, we achieve the
   lowest $\chi^2$ for a $\nu^{-2}$ spectrum which is similar
   to the spectrum of point sources, identified by the WMAP team \cite{mapfor}.
   We should also note that since the $\ell$-dependence of the SZ
   and Point Source signals are very similar, the two signals are
   correlated at a $50-70\%$ level, which is shown in Figure  8. Using a less stringent
   extinction mask ($A_K<0.1$, see IV.B.1) increases the detection
   level of microwave sources by about $10\%$.
\begin{figure}[t]
\includegraphics[width=450pt]{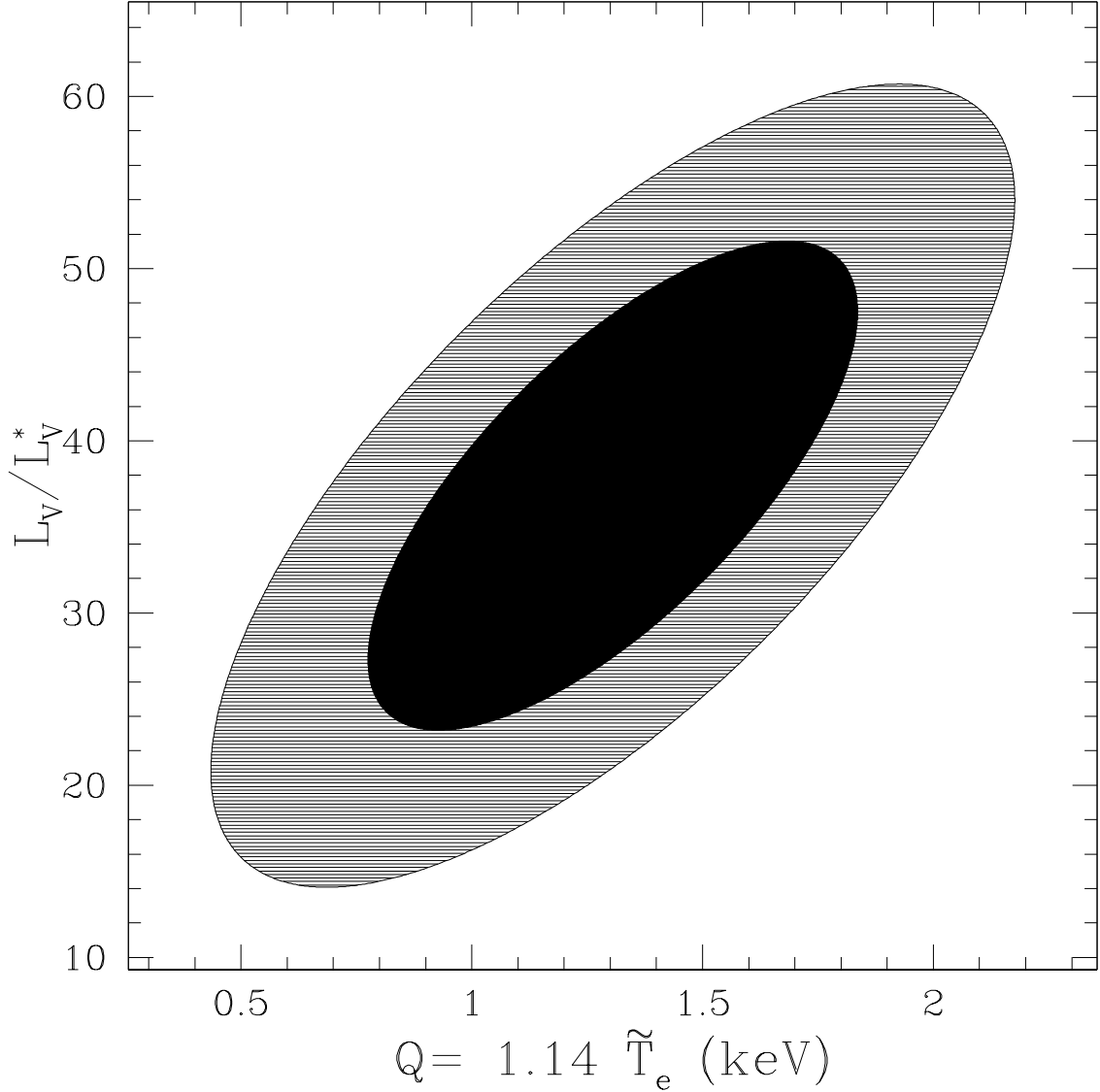}
\caption{1 and $2-\sigma$ likelihood regions of our SZ+Point
Source signals, for a $\delta T_A \propto \nu^{-2}$ spectrum (see
Table II). $Q$ (defined in Eq.A2) is the coefficient of the
mass-temperature relation for galaxy clusters, while $\tilde{T}_e$
 (defined in Eq.7), is the product of gas pressure bias and the average electron temperature.
$L_V$ is the average WMAP V-band luminosity of the 2MASS sources,
while $L^*_V$ (defined in Eq.B6) is the same number, estimated for
the Milky Way. The large correlation of $Q$ (SZ signal) and $L_V$
(Point Source signal) is due to the similar $\ell$-dependence of
the two signals (see Figure 7). Note that the conversion between
$\tilde{T}_e$ and $Q$ depends on the assumed cosmological model
(see Appendix A).}
\end{figure}

     To relax our assumption for the redshift distribution of
     Point Sources (which we assume to be the same as 2MASS sources
     at each magnitude bin; see II.C), we can also allow different
     magnitude bins to have different $L_V$'s and treat each as a free
     parameter. It turns out that this does not affect either our ISW or
     SZ signals, or their significance, while $L_V$'s for each magnitude bin
     is consistent with the values in Table II, within the errors. As the Point Source signal
     is dominated by Poisson noise at large $\ell$'s, removing the assumed
     clustering among the Point Sources (see II.C), does not
     affect the SZ or ISW signals either.

\section{Conclusions}
We obtain the cross-power spectrum of the three highest frequency
bands of the WMAP cosmic microwave background survey with the
2MASS Extended Source Catalog of near infrared galaxies. We detect
an ISW signal at the $\sim 2.5\sigma$ level, which confirms the
presence of a dark energy, at a level consistent with the WMAP
concordance cosmology. We also find evidence for an
anti-correlation at small angles (large $\ell$'s), which we
attribute to thermal SZ. The amplitude is at $3.1-3.7\sigma$ level
and is consistent with the X-ray observations of galaxy clusters.
Finally, we see a signal for microwave Point Sources at the
$2.6\sigma$ level.

We've seen that the completeness limit of the extended source
catalog is between 13.5 and 14 in K.  However, matches with SDSS
show that there are many unresolved sources in the 2MASS Point
Source Catalog (PSC) that are in fact galaxies.  If we can select
out galaxies in the PSC, perhaps by their distinctive colors, we
should be able to push the sample at least half a magnitude
fainter than we have done here, probing higher redshifts with a
substantially larger sample.

Future wide-angle surveys of galaxies should be particularly
valuable for cross-correlation with the WMAP data, especially
as the latter gains signal-to-noise ratio in further data releases.
The Pan-STARRS project \citep{kaiser} for example, should yield a
multi-color galaxy catalog to 25th mag
or even fainter over 20,000 square degrees or more of the sky
well before the end of the decade; it will more directly probe
the redshift range in which the SZ and ISW kernels peak, and
therefore should be particularly valuable for cross-correlating
with WMAP and other CMB experiments.

\acknowledgments NA wishes to thank David N. Spergel for the
supervision of this project and useful discussions. We would also
like to thank Eiichiro Komatsu, Andrey Kravstov and Christopher
Hirata for illuminating discussions and helpful suggestions, Doug
Finkbeiner for help on the analysis of WMAP temperature maps, and
R.M. Cutri and Mike Skrutskie on the 2MASS dataset. MAS
acknowledges the support of NSF grants ASF-0071091 and
AST-0307409.

%\begin{figure}[t]
%\includegraphics[width=450pt]{LRG_ZDist.eps}
%\caption{\label{fig:lrg_dndz}Photometric redshift distributions
%for the four LRG samples, taking into account the covariance
%between photometric redshift and galaxy type.}
%\end{figure}

\def\refe {\par \hangindent=.7cm \hangafter=1 \noindent}
\def\apj { ApJ }
\def\astroph{{\tt astro-ph/}}
\def\aap {A \& A }
\def\ajs{ ApJS }
\def\apss{ Ap\&SS }
\def\aj{AJ}
\def\prd{Phys ReV D}
\def\apjs{ ApJS }
\def\mnras { MNRAS }
\def\apjl { Ap. J. Let. }

\appendix
\section{Semi-Analytical Estimate of SZ Signal}

    In order to find $\tilde{T}_e$ (defined in Eq.7) we need an expression for the
    dependence of the
    electron pressure overdensity on the matter overdensity. As
    the shock-heated gas in clusters of galaxies has keV scale
    temperatures and constitutes about $5-10\%$ of the baryonic
    mass of the universe, its contribution to the average
    pressure of the universe is significantly higher than the photo-ionized
    inter-galactic medium (at temperatures of a few eV).
    Thus, the average electron pressure in a large region of space with average density $\bar{\rho}(1+\delta_m)$
    is given by
    \bea
    \delta p_e \simeq \frac{\bar{n}_e}{\bar{\rho}}\int dM \cdot M
    \cdot k_B
     [T_e(M;\bar{\rho}) \frac{\partial{n(M;\bar{\rho})}}{\partial \bar{\rho}}+ n(M;\bar{\rho})
     \frac{\partial T_e(M;\bar{\rho})}{\partial \bar{\rho}}]
     \bar{\rho} \delta_m \nonumber\\=\frac{\bar{n}_e}{\bar{\rho}}\int dM \cdot M \cdot n(M;\bar{\rho})
     [k_B T_e(M;\bar{\rho})]
     [b(M)+ \frac{\partial \log T_e}{\partial \log \bar{\rho}}] \delta_m,
    \eea
    where $n(M;\bar{\rho})$ and $T_e(M;\bar{\rho})$ are the mass function and
    temperature-mass relation of galaxy clusters
    respectively. Also, $b(M)=\frac{\partial \log n(M;\bar{\rho})}{\partial \log \bar{\rho}}$ is the bias factor
    for haloes of virial mass $M$ ($=M_{200}$; mass within the sphere with the overdensity of 200
    relative to the critical density).
    For our analysis, we use the Sheth \& Tormen analytic form \citep{ShethTormen1999}, for $n(M)$ and
    $b(M)$, which is optimized to fit numerical N-body
    simulations.

    We can use theoretical works on the cluster mass-temperature relation
    (which assume equipartition among thermal and kinetic energies of different components in the intra-cluster medium)
    to find $T_e(M)$, (e.g. \citep{Afs02})
    \bea
    \frac{k_B T_e(M)}{m_p} &\simeq& (0.32 Q) (2\pi G H M)^{2/3}\nonumber\\
    \Rightarrow T_e(M)&=&(6.62 ~\kev)Q \left(\frac{M}{10^{15} h^{-1}
    M_{\odot}}\right)^{2/3},\nonumber\\
    {\rm while}~~&& 1<Q<2
    \eea
      for massive clusters, where $H=100h ~\kms/\mpc$ is the (local) Hubble constant. Although there is controversy on the value of the normalization $Q$
      (see e.g. \citep{eiichiro} and references there in), \citep{Afs02} argue that, as long as there are no
      significant ongoing astrophysical feed-back or cooling (i.e., as long the evolution is adiabatic),
      the dependence on $H$ and $M$ should be the same. Combining this
      with the local comoving continuity equation
      \beq
      3(H+\delta H) = -\frac{\dot{\bar{\rho}}}{\bar{\rho}} -\dot{\delta}_m,
      \eeq
      yields
      \beq
      \frac{\partial \log T_e}{\partial \log
      \bar{\rho}}=\frac{2}{3}\frac{\partial \log H}{\partial \log
      \bar{\rho}} = -\frac{2\dot{D}}{9DH},
      \eeq
      where $D$ is the linear growth factor.

      One may think is that observations may be the most reliable way of constraining $Q$ in Eq.(A2).
      However, almost all the observational signatures of the hot gas in the intra-cluster
      medium come from the X-ray observations which systematically
      choose the regions with high gas density. With this in mind, we should mention that
      while observations prefer a value of $Q$ close to 1.7, numerical
      simulations and analytic estimates prefer values closer to
      $1.2$\citep{Afs02}. For our analysis, we treat $Q$
      as a free parameter which we constrain using our cross-correlation data (see Sec. V.A).

      Putting all the pieces together, we end up with the following expression for $\tilde{T}_e$
      \bea
      &\tilde{T}_e = (0.32~ Q) (2\pi GH)^{2/3}&\nonumber\\ &\times\int d\nu
      f_{ST}[\nu] M^{2/3} [b_{ST}(\nu)-\frac{2\dot{D}}{9DH}],& \\
      &\nu(M)= \frac{\delta_c}{\sigma(M)}&\nonumber,
      \eea
      where $\sigma(M)$ is the variance of linear mass overdensity
      within a sphere that contains mass $M$ of unperturbed
      density, while $\delta_c \simeq 1.68$ is the spherical
      top-hot linear growth threshold\citep{gunn}. $f_{ST}$ and
      $b_{ST}$ are defined in \citep{ShethTormen1999}. For the
      WMAP concordance cosmological model\citep{BennettEtAl2003a}, this integral can be
      evaluated to give
      \beq
      \tilde{T}_e = b_p \bar{T}_e = (0.88 ~ Q) ~\kev.
      \eeq

        The above simple treatment of the SZ signal fails at scales
      comparable to the minimum distance between clusters, where the average gas
      pressure does not follow the average matter density
      \citep{refreiger,zhang}, which leads to a scale-dependent
      pressure bias. Moreover, efficient galaxy formation removes
      the hot gas from the intra-cluster medium, which causes
      Eq.A1 to overestimate the SZ signal. As this paper mainly
      focuses on the observational aspects of our detection, we
      delay addressing these issues into a further
      publication\citep{komafs}. The preliminary results seem to
      be consistent with the above simple treatment at the
      20$\%$ level.

\section{Microwave Luminosities of the Andromeda Galaxy and the Milky Way }

      First we derive how much the flux received by a microwave
      source at distance $d_L$ and observed solid angle $\delta \Omega$ affects the observed CMB
      temperature. The apparent change in the black-body
      temperature is obtained by
      \beq
      \delta \Omega ~\cdot \delta\left[ \frac{4\pi(\hbar/c^2) \nu^3 \Delta
      \nu}{\exp[h\nu/(k_B T_{\cmb})]-1}\right] = \frac{L}{4\pi d^2_L}.
      \eeq
        The left hand side of Eq.(B1) is the change in the Planck
        intensity, where $\nu$ and $\Delta \nu$ are the detector
        frequency and band width respectively. The right hand side
        is the observed Microwave flux. Defining $x$ as the
        frequency in units of $k_B T_{\cmb}/h$, Eq.(B1) yields
      \beq
       \frac{\delta T}{T} = \frac{4\pi^2 \hbar^3 c^2}{(k_B T_{\cmb})^4}\cdot\frac{\sinh^2(x/2)}{x^4 \Delta x}\cdot
       \frac{L}{\delta \Omega d^2_L}.
      \eeq

         To obtain the microwave luminosity of Milky Way, we
         assume an optically and geometrically thin disk, with
         a microwave emissivity, $\epsilon$, which is constant across its
         thickness and falls as $\epsilon_0 \exp(-r/r_0)$ with the distance, $r$, from
         its center. The disk thickness is $2H \ll r$, while we assume
         $r_0 \simeq 5 \kpc$, our distance from the Galactic center
         is $r \simeq 8.5 \kpc$, and our vertical distance from
         the center of the disk is $z$. Integrating Eq.(B2) over the disk
         thickness leads to the cosecant law for the Galactic
         emission
       \beq
        \frac{\delta T}{T} (b;r) = \frac{4\pi^2 \hbar^3 c^2}{(k_B T_{\cmb})^4}\cdot\frac{\sinh^2(x/2)}{x^4 \Delta x}\cdot
        \epsilon_0 e^{-r/r_0} (H|\csc b| - z \csc b),
       \eeq
         where $b$ is Galactic latitude. Integrating $\epsilon(r)$ over the disk volume gives the total luminosity of
         the Milky Way
       \beq
         L = 2H \int~ 2\pi r dr~ \epsilon_0 e^{-r/r_0} = 4\pi H r^2_0 \epsilon_0.
       \eeq
         Combining Eqs.B3 and B4, we can obtain the total luminosity
         of Milky Way from the observed Galactic emission
       \beq
         L = r_0^2 e^{r/r_0} \cdot \frac{(k_B T_{\cmb})^4}{2\pi
         \hbar^3 c^2} \cdot \frac{ x^4 \Delta x}{\sinh^2 (x/2)}
         \cdot |\sin b| \left[\frac{\delta T}{T}(b;r) +\frac{\delta
         T}{T}(-b;r)\right].
       \eeq

         Figure 7 in \citep{mapfor} gives the cosecant law for the
         Galactic emission in different WMAP bands. Using this
         information in Eq.(B5) (after conversion into thermodynamic units) gives the luminosity of the Milky
         Way in WMAP bands
       \bea &L^*_Q = 1.7 \times 10^{37} ~\ergs,&\nonumber\\
~ &L^*_V = 3.0 \times 10^{37} ~\ergs,&\nonumber\\ &{\rm and}~
L^*_W = 1.0 \times 10^{38} ~\ergs. &
 \eea

         To confirm these values, we can use Eq.(B2) and the observed integrated flux of
         the Andromeda (M31) galaxy in the WMAP maps to obtain its microwave
         luminosity
\bea &L_{M31,Q} = 2.1 \times 10^{37} ~\ergs,&\nonumber\\
~ &L_{M31,V} = 5.3 \times 10^{37} ~\ergs,&\nonumber\\ &{\rm and}~
L_{M31,W} = 1.6 \times 10^{38} ~\ergs. &
 \eea
         We see that these values are larger, but within
         $50\%$, of the Milky Way microwave luminosities.

\begin{thebibliography}{30}
\expandafter\ifx\csname
natexlab\endcsname\relax\def\natexlab#1{#1}\fi

\bibitem{ACO}
Abell, G.O., Corwin, H., Olwin, R., 1989, ApJS, 70, 1

\bibitem[Afshordi \& Cen(2002)]{Afs02}
Afshordi, N., \& Cen, R. 2002, ApJ, 564, 669

\bibitem[Bell \etal(2003)]{Bel03}
Bell, E.F., \etal 2003, ApJS, 149, 289

\bibitem{cobe}
Bennett, C. \etal 1996, ApJ, 464, L1

\bibitem[Bennett {\it et~al.} (2003)]{BennettEtAl2003a}
Bennett,~C.L. {\it et~al.} 2003, ApJS, 148, 1; The public data and
other WMAP papers are available at
http://lambda.gsfc.nasa.gov/product/map

\bibitem[Bennett {\it et~al.} (2003)]{mapfor}
Bennett,~C.L. {\it et~al.} 2003, ApJS, 148, 97

\bibitem{Steve}
Boughn, S.~P., Private Communication.

\bibitem[Boughn \& Crittenden(2002)]{Bou02}
Boughn, S.~P. \& {Crittenden}, R.~G., Phys. Rev. Lett. 88, 021302
(2002)

\bibitem[{{Boughn} \& {Crittenden}(2003)}]{boughn/crittenden:2003}
Boughn, S.~P. \& {Crittenden}, R.~G. 2003, astro-ph/0305001

\bibitem{nvss}
Condon,~J. \etal 1998, Astron. J. 115, 1693

\bibitem[Cooray (2002)]{Cooray2002}
Cooray,~A. 2002, Phys. Rev. D, 65, 103510

\bibitem[Corasaniti {\it et al.} (2003)]{CorasanitiEtAl2003}
Corasaniti,~P.S. {\it et al.} 2003, PRL, 90, 091303

\bibitem{Turok}
Crittenden, R.G., \& Turok, N. 1996, PRL, 76, 575

\bibitem[Diego, Silk \& Sliwa (2003)]{diego}
Diego.~J.M., Silk~J., \& Sliwa 2003, MNRAS, 346, 940

\bibitem{efstathiou}
Efstathiou, G. 2003, astro-ph/0307515

\bibitem[{Fosalba} \& {Gazta\~{n}aga} (2003)]{fosalba1}
Fosalba P., \& Gazta\~{n}aga E. 2003, astro-ph/0305468

\bibitem[{Fosalba}, {Gazta\~{n}aga} \& Castander (2003)]{fosalba2}
Fosalba P., Gazta\~{n}aga E., \& Castander, F.J. 2003, ApJ, 597L,
89

%\bibitem{loeb}
%Fox, D.C., \& Loeb, A. 1997, ApJ, 491, 459

\bibitem{healpix}
Gorski, K. M., Hivon, E., \& Wandelt, B. D. 1998, in
\emph{Evolution of Large-Scale Structure: From Recombination to Garching}

\bibitem{gunn}
Gunn, J., Gott, J. 1972, ApJ, 176, 1

\bibitem{hernand}
Hernandez-Monteagudo, C., \& Rubino-Martin, J.A. 2003,
astro-ph/0305606

\bibitem{hivon}
Hivon, E., Gorski, K. M., Netterfield, C. B., Crill, B. P.,
Prunet, S., \& Hansen, F. 2002, ApJ, 567, 2

\bibitem{hinshaw}
Hinshaw, G., \etal 2003, ApJS, 148, 135

\bibitem[Hu \& Dodelson (2002)]{HuDodelson2002}
Hu,~W. \& Dodelson,~S. 2002, ARA\&A, 40, 171

\bibitem[Huchra \& Mader(2001)]{Huc01}
Huchra, J \& Mader, J (2000) at
{\rm http://cfa-www.harvard.edu/$\sim$huchra/2mass/verify.htm}


\bibitem[Ivezic \etal(2000)]{Ive00}
Ivezi${\rm \acute{c}}$, ${\rm \check{Z}}$. \etal in
\emph{IAU Colloquium 184: AGN Surveys, 18-22 June 2001, Byurakan(Armenia)}

\bibitem[Jarrett \etal (2000)]{Jar00}
Jarrett, T.H., \etal 2000, AJ, 119, 2498

\bibitem{kaiser}
Kaiser et al. 2002, SPIE, 4836, 154

\bibitem{kamion}
Kesden, K., Kamionkowski, M., \& Cooray, A. 2003, astro-ph/0306597

\bibitem[Kochanek \etal (2001)]{Koc01}
Kochanek, C.S., \etal 2001, ApJ 560, 566

\bibitem[Kogut {\it et~al.} (2003)]{KogutEtAl2003a}
Kogut,~ C.L. {\it et~al.} 2003a, ApJS, 148, 161

\bibitem{komafs}
Komatsu, E., Afshordi, N., \& Seljak, U. 2003, in preparation

\bibitem{eiichiro}
Komatsu, E., \& Seljak, U. 2001, MNRAS, 327, 1353

\bibitem{limber}
Limber, ~D.N. 1954, \apj, 119, 655

\bibitem{apm}
Maddox, S. J., et al. 1990, MNRAS, 242, 43

\bibitem[Maller \etal (2003)]{Mal03}
Maller, A. H., \etal 2003, astro-ph/0304005

\bibitem{Myers2003}
Myers, A.D., Shanks, T. Outram, P.J., Wolfendale, A.W. 2003,
astro-ph/0306180

\bibitem{Nikolaev2000}
Nikolaev, S., \etal 2000, AJ, 120, 3340

%\bibitem{Nichol2003}
%Nichol, R.C. {\it et al.} 2003, in preparation


\bibitem[Nolta \etal (2003)]{Nol03}
Nolta, M.R., \etal 2003, astro-ph/030597
\bibitem{peacock}
Peacock, J.A., \& Dodds, S.J. 1996, MNRAS, 280L, 19

\bibitem{ratra}
Peebles, P. J.~E., \& Ratra, B. 2003, Rev.Mod.Phys. 75, 599

\bibitem[Peiris \& Spergel (2000)]{PeirisSpergel2000}
Peiris,~H., \& Spergel,~D.N. 2000, ApJ, 540, 605

\bibitem{refreiger}
Refregier A., Komatsu E., Spergel D. N., Pen U., 2000, Phys. Rev.
D, 61,123001

\bibitem[Sachs \& Wolfe (1967)]{SachsWolfe1967}
Sachs,~R.~K. \& Wolfe,~A.~M. 1967, ApJ, 147, 73

\bibitem[Schechter(1976)]{Sch76}
Schechter, P. 1976, ApJ 203, 296.

\bibitem[Schlegel, Finkbeiner \& Davis(1998)]{Sch98}
Schlegel, D.J., Finkbeiner, D.P. \& Davis, M. 1998, ApJ 500, 525.

\bibitem[Scranton et al.(2003)]{scranton}
Scranton, R.~et al. 2003, astro-ph/0305337

\bibitem{cmbfast}
Seljak, U. \& Zaldarriaga, M. 1996, ApJ, 469, 437

\bibitem[Sheth \& Tormen (1999)]{ShethTormen1999}
Sheth,~R.~K. \& Tormen,~G. 1999, MNRAS, 308, 119

\bibitem[Spergel {\it et al.} (2003)]{SpergelEtAl2003}
Spergel,~D.N. {\it et~al.} 2003, ApJS, 148, 175

\bibitem[Skrutskie \etal (1997)]{Skr97}
Skrutskie \etal 1997, in \emph{The Impact of Large Scale Near-IR Sky Survey},
ed. F. Garzon \etal (Dordrecht: Kluwer), 187

\bibitem{tegmark}
Tegmark, M. 1997, Phys. Rev. D, 56, 4514

\bibitem[York {\it et al.} (2000)]{YorkEtAl2000}
York,~D.G. {\it et al.} 2000, AJ, 120, 1579

\bibitem{zhang}
Zhang, P., \& Pen, U. 2001, ApJ, 549, 18

\bibitem[Zel'dovich \& Sunyaev (1967)]{SZ}
Zel'dovich~Y.B., \& Sunyaev,~R.A. 1969, ApSpSci, 4, 129.

\end{thebibliography}
\end{document}